\documentclass[9pt,twocolumn,twoside]{osajnl}

\journal{ao} % Choose journal (ao, aop, josaa, josab, ol)
\usepackage{xspace}
\usepackage[utf8]{inputenc} 
\makeatletter
\newcommand{\MathFuncName}[1]{{\operator@font #1}}
\newcommand{\MathFunc}[1]{\mathop{\operator@font #1}\nolimits}
\newcommand{\MathFuncWithLimits}[1]{\mathop{\operator@font #1}\limits}
\makeatother

\newcommand{\mathd}{\mathrm{d}}          % roman 'd' for integrand/differential
\newcommand{\V}[1]{\boldsymbol{#1}}      % vector
\newcommand{\M}[1]{\mathbf{#1}}          % matrix
\newcommand{\TransposeLetter}{\top}
\newcommand{\T}{^{\TransposeLetter}}     % suffix for transpose
\newcommand{\D}[1]{{\mathd #1}}          % integrand/differential
\newcommand{\Var}{\MathFunc{Var}}        % variance
\newcommand{\prox}{\operatorname{prox}}  % proximity operator
\newcommand{\proxy}[1]{\widetilde{#1}}   % approximation
\newcommand{\Eq}[1]{(\ref{#1})}
\newcommand{\Fig}[1]{Figure~\ref{#1}}

\newcommand{\YORICK}[1]{}

\newcommand{\bigParen}[1]{\bigl(#1\bigr)}

\newcommand{\norm}[1]{\Vert #1\Vert}
\newcommand{\Norm}[1]{\left\Vert #1\right\Vert}

\newcommand{\abs}[1]{\vert #1\vert}
\newcommand{\Abs}[1]{\left\vert #1\right\vert}

\newcommand{\SNR}[1]{\textrm{SNR}(#1)}
\newcommand{\SNRx}{\SNR{\V{x}}}
\newcommand{\SNRxb}{\SNR{\V{x}^\best}}

\newcommand{\argmin}{\MathFuncWithLimits{arg\,min}}

\newcommand{\minimize}{\MathFuncWithLimits{minimize}}
\newcommand{\find}{\MathFuncWithLimits{find}}

\newcommand{\Reals}{\mathbb{R}}

\newcommand{\Complexes}{\mathbb{C}}

\newcommand{\bydef}{\stackrel{\text{\tiny def}}{=}}

\newcommand{\eg}{\emph{e.g.}\xspace}
\newcommand{\ie}{\emph{i.e.}\xspace}

\newcommand{\micron}{\text{\textmu{}m}} % requires textcomp
\newcommand{\best}{+}

\graphicspath{{figs/}} 

\title{Proximity Operators for Phase Retrieval}

\author[1,*]{Ferreol Soulez} \author[2]{\'Eric Thiebaut} \author[3]{Antony
Schutz} \author[3]{Andre Ferrari} \author[4]{Fredeic Courbin}
\author[1]{Michael Unser}

\affil[1]{Biomedical Imaging Group, \'Ecole polytechnique federale de Lausanne
(EPFL), Lausanne CH-1015, Switzerland. } 
 \affil[2]{Univ Lyon, Univ Lyon1, Ens de Lyon, CNRS, Centre de Recherche Astrophysique de Lyon UMR5574, F-69230, Saint-Genis-Laval, France}
\affil[3]{Lab.   J.-L.  Lagrange, Universite de  Nice  Sophia  Antipolis,
CNRS, Observatoire  de  la  Côte  d'Azur,  Parc  Valrose,  F-06108  Nice  cedex
02,  France}
\affil[4]{Laboratoire d'astrophysique, \'Ecole Polytechnique Federale de
Lausanne (EPFL), Observatoire de Sauverny
CH-1290 Versoix, Switzerland.}

\affil[*]{Corresponding author: ferreol.soulez@epfl.ch}

\dates{Compiled \today}

\ociscodes{100.5070, 100.3020, 100.3190,100.6640}

\begin{abstract}
We present a new formulation of a family of  proximity operators that
generalize the projector step for phase retrieval. These proximity operators for
noisy intensity measurements  can replace the classical
``noise free'' projection  in any projection-based algorithm. They are
derived from a \emph{maximum likelihood} formulation and admit  closed form
solutions for both the Gaussian and the Poisson cases. In addition, we extend
these proximity operators to undersampled intensity measurements.
To assess
their performance, these operators are exploited in a classical Gerchberg Saxton
algorithm. We present numerical experiments showing that the reconstructed
complex amplitudes with these proximity operators perform always  better than
using the classical intensity projector while their computational overhead is
moderate.
\end{abstract}

\setboolean{displaycopyright}{true}

\begin{document}

\maketitle 

\thispagestyle{fancy}

%\ifthenelse{\boolean{shortarticle}}{\ifthenelse{\boolean{singlecolumn}}{\abscontentformatted}{\abscontent}}{}

\section{Introduction}
\label{sec:intro}
%-- ML simpler than MAP
%-- GS -> ML

The classical phase-retrieval problem is to reconstruct a complex-valued signal
$\V{x}$ from measurements of its squared modulus  \cite{Walther1963}.
This problem arises in many applications
(\eg{}, cristallography \cite{Harrison1993}, microscopy \cite{Misell1973},
astronomy \cite{FienupDainty1987}). Since the seminal paper of
Gerchberg and Saxton \cite{Gerchberg1972}, an abundant literature has been
devoted to it (see \cite{ShechtmanEldarCohenEtAl2015} for a review).
A large part of
the proposed algorithms relies on successive projections
\cite{Fienup1980,Fienup1982,Elser2003,BauschkeCombettesLuke2003,Luke2005}.
For the last few years, there is a renewed interest for phase retrieval
and several new alternatives to successive projections methods have been
proposed:
the semi-definite-programming based formulations \cite{Candes2013,FogelWaldspurger2013}, the algorithms for phase retrieval of sparse signal
\cite{Shechtman2014,CaiLiMa2015,TillmannEldarMairal2016}
the gradient based methods using Wirtinger derivatives
\cite{CandesLiSoltanolkotabi2015,ChenCandes2015} and a variational
Bayesian framework \cite{DremeauKrzakala2015}.

Here, we adopt a vector representation of the complex image $\V{x} =
(x_1,\dots,x_K)$, where $K$ is the number of pixels. In the phase retrieval  
problem, the forward model that links the complex amplitude $\V{x} \in
\Complexes^K $ to  the measured  image intensities $\V{d}\in \Reals^{K}_{+} $
is
\begin{equation}
d_k = \Abs{x_k}^2 + n_k\,,
\end{equation}
where $\V{n}$ is some measurement noise and $\Abs{x_k}^2$ denotes the squared
modulus of $x_k$.

Such an inverse problem is classically solved in a variational framework by
  estimating $\V{x}$ that  minimizes a cost function
\begin{equation}
%\label{costfunction}
\label{eq:invpb}
 \mathcal{C}(\V{x}) =\mathcal{L}(\V{x}) + \mathcal{R}(\V{x})\,,
\end{equation}
  which is a sum of the data term
$\mathcal{L}$ and a regularization functional $\mathcal{R}$.
In this approach known as penalized \emph{maximum likelihood}, the data term is
defined according to the forward model and the statistics of the noise, 
whereas the regularization function is designed to enforce some prior knowledge
about $\V{x}$ (such as support, non-negativity, smoothness,\dots). As
$\mathcal{L}$  and $\mathcal{R}$ are defined
independently, any improvement on one of these functions implies a better
estimate of the solution of the inverse problem.

Most projection-based algorithms
\cite{Gerchberg1972,Fienup1980,Fienup1982,Elser2003,BauschkeCombettesLuke2003,Luke2005}
use constraints that assume noise-free measurements. Some
authors have studied the behavior of these methods in noisy environment
\cite{WilliamsPfeiferVartanyantsEtAl2007}  while others have  proposed
 empirical modifications to mitigate the effect of the noise 
\cite{LatychevskaiaLongchampFink2011,DilanianWilliamsWhiteheadEtAl2010,MartinWangLohEtAl2012}.
In this paper, we derive a likelihood function adapted to the statistics of
the noise via a simple  modification of the intensity-projection operator.
We had previously established the formulation of  this proximity operator in the
Gaussian case with a specific ADMM algorithm for image reconstruction in optical
long-baseline interferometry \cite{Schutz2014PAINTER} ; a similar result was
also published recently \cite{WellerPnueliDivonEtAl2015}.
But neither further characterization nor comparison with standard projection
methods were done. 
 
% At the difference of \cite{MartinWangLohEtAl2012}, our projection operator for
% noisy intensity can be plugged in any projection based algorithm without
% changing the core of the optimization.
Rather than a full-fledged phase-retrieval algorithm, the scope of this paper is
% therefore not to propose a new phase retrieval algorithm but rather
a novel formulation of a noise-adapted projection step that can be used in any
other projection-based algorithm \cite{Fienup1982}.
Therefore, we  focus  on the likelihood function.
To demonstrate its effect, we  apply the proposed projectors in the standard
Gerchberg Saxton algorithm (GS).
We have chosen this simple phase-retrieval algorithm as a baseline because it
does not use any priors.
Therefore, the quality of its results  depend only on the projection used. %This
% will highlight the effect of our proximity of operators.
Whereas  GS is hardly state-of-the-art anymore, the reader must keep in mind
that our proposed proximity operators  can be plugged into many optimization
schemes (see \cite{BauschkeCombettesLuke2003,CombettesPesquet2011}) that rely
on proximity operators to minimize a regularized cost function.

\section{Gerchberg-Saxton algorithm}

The error-reduction method  (GS), described in
Algorithm~(\ref{alg:GS}),  estimates the complex amplitude (the
wavefront) of a light wave in the plane $z_A$ from the  intensity profiles 
$\V{d}_A$ and $\V{d}_B$ measured at  depth $z_A$ and $z_B$, respectively.  It
solves the feasibility problem
\begin{equation}\label{eq:GSfind}
\find  \V{x}\in C_A \cap \left\{\V{x}:\, \M{H}\,\V{x} \in
C_B \right\}\,,
\end{equation}
% \begin{equation}\label{eq:GSfind}
% \V{x}^\best = \find \V{x}\in \left\{ \V{x}\in C_A \cap \M{H}\,\V{x} \in
% C_B \right\}\,,
% \end{equation}
where  $\M{H}$ is
the propagation operator from plane $z_A$ to plane $z_B$ and $C_i$ with $i={A,
B}$ is the set of complex-valued signals of squared modulus $\V{d}_i$, \ie
$C_i =\{ \V{x} \in \Complexes^K, \Abs{\V{x}}^2 = \V{d}_i \} $. The propagation operator is
classically either the Fresnel operator (under a Fresnel approximation) or the
Fourier operator (under a Fraunhoffer approximation). % with $\M{H}^{-1} =
% \M{H}^{*}$.
 This can be
reformulated as the minimization problem
\begin{equation}
\V{x}^\best \in \argmin_{\V{x}\in \Complexes^K} \left( \imath_{C_A}(\V{x})+
\imath_{C_B}(\M{H}\,\V{x})\right)\,,\label{eq:GSind}
\end{equation}
where $\imath_C$ is the indicator
function of the set $C$ defined as
\begin{equation}
\imath_C(\V{x}) = \left\{
\begin{array}{ll}
0\,, & \text{if } \V{x} \in C\\
+\infty \,, & \text{otherwise.}\\
\end{array}\right.
\end{equation} 
Observe that, when $\V{d}_i \neq \V{0}$, 
$C_i$ is generally not a convex set.
Therefore, only  local convergence can be established \cite{NollRondepierre2015}.

\begin{algorithm}
\caption{Gerchberg-Saxton algorithm}\label{alg:GS}
\begin{algorithmic}[1]
\Procedure{GS}{$\V{d}_A$, $\V{d}_B$}
\State $\V{x}^{(0)} = \sqrt{\V{d}_A}$ \Comment{Initialization}
\For{$n = 1, 2, \dots, \textrm{maxiter}$}
\State $\V{y}^{(n-1/2)} = \M{H}\cdot\V{x}^{(n-1)}$ \Comment{Propagation to the
$z_B$ plane}
\State $\V{y}^{(n)} = P_B(\V{y}^{(n-1/2)})$ \Comment{Projection}
\State $\V{x}^{(n-1/2)} = \M{H}^{-1}\cdot\V{y}^{(n)}$ \Comment{Back
propagation to the $z_A$ plane}
\State $\V{x}^{(n)} = P_A(\V{x}^{(n-1/2)})$  \Comment{Projection}
\EndFor
\State \textbf{return} $\V{x}^{(\textrm{maxiter})}$\Comment{The complex
amplitude in the $z_A$ plane}
\EndProcedure
\end{algorithmic}
\end{algorithm}

The GS algorithm and its successors
\cite{Fienup1982,BauschkeCombettesLuke2003,Luke2005}
involve an element-wise projection operator $P(\V{x}\,|\,\V{d}) =
\bigParen{P(x_1\,|d_1), \dots,P(x_K\,|d_K)}$ that constrains the
modulus of the current iterate $\V{x}$ to be equal to the square root of its
measurement $\sqrt{\V{d}}$ while keeping its phase untouched, as in
\begin{equation}
\label{eq:StrictConstraint}
P(x_k\,|d_k) = \left\{
\begin{array}{ll}
\frac{x_k}{\Abs{x_k} }\sqrt{d_k}\,, & \text{if } \Abs{x_k}  > 0\\
\sqrt{d_k}\,, & \text{otherwise.}\\
\end{array}\right.
\end{equation}
The projection  $P(\V{x}\,|\,\V{d})$ of $\V{x}$ onto the set $C$ of all signals
 of intensity (or squared modulus)
$\V{d}$ will be called ``classical projection'' throughout this
paper. It is a solution of
\begin{equation}
\label{StrictProx}
\minimize_{\V{y}\in \Complexes^N} \left( \imath_C(\V{y}) +
\frac{1}{2}\Norm{\V{x}-\V{y}}^2\right)\,.
\end{equation}
To prevent stagnation of the GS algorithm,  a relaxed projection step $P'$ was 
proposed \cite{LyonDorbandHollis1997,LukeBurkeLyon2002}:
\begin{equation} 
\label{eq:MITA}
P'(x_k\,|d_k) =  (1- \beta)\,x_k + \beta \,P(x_k\,|d_k)\,,
\end{equation}
where  $0\le\beta\le 1$ is a relaxation parameter empirically set close to $0$
for regions where the noise dominates.

As observed by Levi and Stark \cite{LeviStark1984,BauschkeCombettesLuke2002},
the  GS algorithm is a non-convex instance of the projection-onto-convex-set
(POCS) algorithm.
%  as the
% modulus constraint \Eq{eq:StrictConstraint} is not convex estimating the
% solution $\V{x}^\best$ such  $\Abs{\V{x}^\best}^2 = \V{d} $.
POCS is
widely employed in signal processing to solve feasibility problems. However, as
soon as noisy intensities are considered, equation
(\ref{eq:StrictConstraint}) does not anymore give  the  solution that is optimal
in the maximum-likelihood sense. Therefore,  GS leads to errors in the
reconstructed wavefront in the presence of noisy measurements.
% Ignoring the fact that intensities are
% intrisicaly corrupted by some measurement noise $n$,  using

We assume that the measurement noise $n_k =     d_k - \Abs{x_k}^2$ at pixel $k$ 
is independent and centered with a probability density $\Pr\left(n_k\,
|\, x_k \right) $.
For a given intensity measurement $d_k$, the co-log-likelihood of  the noise
distribution at pixel $k$ (up to the constant $\textrm{cst}$) is:
\begin{equation}
\ell_k(n_k)= - \ln \Pr \left(n_k\,|\, x_k\right) + \textrm{cst}
\label{eq:lkldef}\,.
\end{equation}
The problem addressed by GS % (Eq. \ref{eq:GSfind})
has a maximum-likelihood formulation expressed by
\begin{equation}
\label{eq:I-lkl}
\V{x}^\best \in \argmin_{\V{x}\in\Complexes^K} \left(\sum_{k=1}^{K}
\ell_k\left(\Abs{x_k}^2 -  d_k\right) + \sum_{k'=1}^{K}
\ell_{k'} \left(\Abs{\left[\M{H}\,\V{x}\right]_{k'}}^2 - d_{k'}\right)\right)\,.
\end{equation}
This is not  a feasibility problem anymore. However, it is still closely related
to the GS formulation described by Equation (\ref{eq:GSfind}).   We  argue
that, with the help of proximal operators, both problems can be solved using
identical convex-optimization techniques (\eg{}, Douglas-Rachford) without
relying on smooth approximations of $\ell$
\cite{RepettiChouzenouxPesquet2014}.

\section{Proximity operator for intensity}
\label{sec:prox}
\subsection{Non-Convex Proximity Operators}
It is possible to tackle a class of problems broader than feasibility problems
by introducing proximity operators \cite{CombettesPesquet2011}.  A proximity
operator (or Moreau proximal mapping \cite{Moreau1965}) is a generalization of
the classical projection  on a set where the indicator  function
$\imath_C$ in \Eq{StrictProx} is replaced by an arbitrary lower
semi-continuous convex function $ g:  \Complexes^K \to \Reals$ so that
\begin{equation}
\label{eq:prox-def}
\prox_{ g}(\proxy{\V{x}})
\bydef \argmin_{\V{x} \in \Complexes^K}
\left( g(\V{x}) + \frac{1}{2}\,\Norm{\V{x} - \proxy{\V{x}}}_2^2\right)
\, .
\end{equation}

The concept of proximal mapping has also been extended to non-convex functions
that fulfill three conditions: (i) lower semi-continuity; (ii)
\emph{prox-boundedness}; and (iii) \emph{prox-regularity} (see Theorem~4 of
\cite{HareSagastizabal2009}).

%\subsection{Analytic formulation}
\subsection{Proximal Operator for Maximum-Likelihood}
\label{sec:proxlkl}

%\section{Maximum likelihood solutions}

%  The log-likelihood function reads:
% \begin{equation}
% \lkl(\V{x}) = \sum_{k=1}^N \ln
% f_k(x_k)\,.
% \end{equation}

\begin{figure}[tbp]
\noindent\begin{minipage}[t]{.45\linewidth}
\fbox{\includegraphics[width=\linewidth]{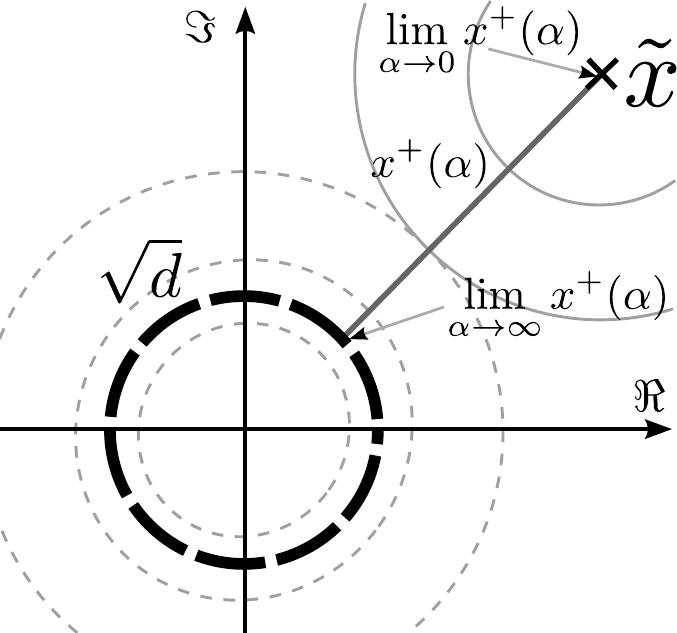}}
\caption{Trajectory of $x^\best(\alpha) = \prox_{\alpha
f}(\proxy{x})$ as a function of $\alpha$. $x^\best(\alpha)$ follows the line
where the level set of $f$ (thin dashed circles) and $\Abs{x - \proxy{x}}^2$
(thin circles) are tangent.
}
\label{fig:PwrSpctr}
\end{minipage}\hfill
\begin{minipage}[t]{.45\linewidth}
\fbox{\includegraphics[width=\linewidth]{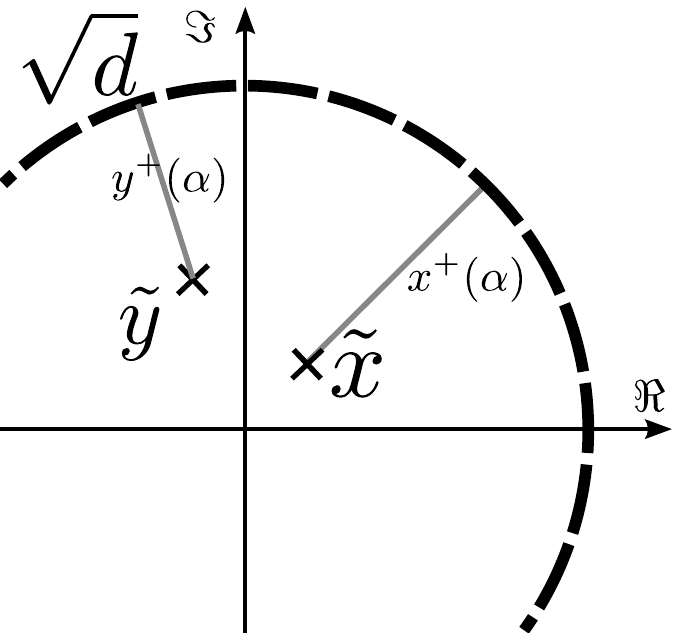}}
\caption{ Illustration of the expansiveness of $\prox_{\alpha
f}(x)$ when $\Abs{x}^2 < d$}.
\label{fig:PwrSpctr2}
\end{minipage}
\end{figure}

As long as the measurement noise is uncorrelated, the  likelihood function
defined in \Eq{eq:lkldef} is separable along pixels. 
In this element-wise operation,  we shall drop the
subscript $k$ to simplify the notations and state $\alpha\,f(x) =
\ell_k(d_{k} - \abs{x_k}^2)$ with  $\alpha>0$ a tuning factor.
The function $f$ has the following properties: (i)   \emph{continuity}, provided
that $\ell_k$ is also continuous (that is true for most
noise statistics used in practice) ; (ii)
 \emph{non convexity} (\eg{}, if 
$x_1= \sqrt{d}$ is a minimum of $f$, then $x_2=-\sqrt{d}$ is, but not
necessarily  $(x_1 + x_2)/2$) ; (iii) \emph{prox-boundedness}  as $f$ is
 positive (and proper).
%(ii) \textbf{continuity} ($f$ is $C^\infty$),
%(iii)  
However, as described further,   $f$ is not  \emph{prox-regular} in $x = 0$.

The proximity operator of $\alpha\, f$  is given by
\begin{equation}
\label{eq:proxI-def}
\prox_{\alpha f}(\proxy{x})
= \argmin_{x \in \Complexes} \,\left\{ \alpha\,f (x) +
\frac{1}{2}\,\Abs{x - \proxy{x}}^2 \right\}\, .
\end{equation}
% \begin{equation}
%   \label{eq:proxI-def}
%   \prox_{\alpha f_k}(\proxy{x_k})
%   = \argmin_{x \in \Complexes} \alpha \, w_k (\Abs{x}^2 -
% d_k)^2 + \frac{1}{2}\,\Abs{x -
%   \proxy{x_k}}^2 \, .
% \end{equation}

As $f$ is  a function that depends only on  the squared modulus of $x$, the
solution necessarily lies on the line passing through $\proxy{x}$ and $0$ where the
gradients of both  parts of \Eq{eq:proxI-def} have opposite directions.
The phase of the solution is therefore  the phase of $\proxy{x}$. The solution 
$x^\best = \rho^\best \exp(\jmath\,\phi^\best)$ of \Eq{eq:proxI-def} is
 given by
\begin{eqnarray}
% \rho^\best &=& \argmin_{\rho\ge 0} \alpha\,  w
% (\rho^2 - I)^2 + \frac{1}{2}\,(\rho -
%   \proxy{\rho})^2 \, ,\label{eq:rho}\\
\rho^\best &=& \argmin_{\rho\ge 0} \, \left\{ \alpha\,f
(\rho) + \frac{1}{2}\,(\rho -
\proxy{\rho})^2 \right\}\, ,\label{eq:rho}\\
\phi^\best &=&  \proxy{\phi}\,,
\end{eqnarray}
where $ \proxy{x} =   \proxy{\rho} \exp(\jmath\, \proxy{\phi})$.

Furthermore, if $f(x)$ has its minimum in $\Abs{x}^2=d$ and $f(\Abs{x})$
increases monotonically for $\Abs{x}>\sqrt{d}$, then there is a solution that
lies on the line between $\proxy{x}$ and its projection on the circle $\Abs{x}^2
= d$,  as illustrated  in Figure~\ref{fig:PwrSpctr}. The  position on this line
varies monotonically with  $\alpha$, so that  $\prox_{\alpha
f}(\proxy{x})$ is $\proxy{x}$ for $\alpha=0$ and gets closer to
$\frac{\proxy{x}}{\Abs{\proxy{x}}}\sqrt{d_k}$  as $\alpha$ increases.  The
classical operator defined in \Eq{eq:StrictConstraint} can thus be seen as
$\lim_{\alpha \to \infty} \prox_{\alpha f}(\proxy{x}) = p(x_k\,|d_k) $.
From this solution, we can identify three subdomains where $\prox_{\alpha
f}(\proxy{x})$  has different properties.
\begin{itemize}
  \item When $\proxy{x} \in \{ x
  \in \Complexes , \Abs{x}^2 \ge d \} $, $\prox_{\alpha
  f}(\proxy{x})$  is single valued and thus
  $f$ is \emph{prox-regular}. Furthermore, the proximity operator of $f$ is
  non-expansive on this sub-domain.
  \item When   $\proxy{x} \in \{ x
  \in \Complexes , 0<\Abs{x}^2 < d \} $, $f$ is still  \emph{prox-regular}
  but $\prox_{\alpha
  f}$ is no longer non-expansive. Indeed, $\norm{\prox_{\alpha f}(\proxy{x}) -
  \prox_{\alpha f}(\proxy{y})}_2^2 \ge \norm{\proxy{x} - \proxy{y}}_2^2$, as
  illustrated  in \Fig{fig:PwrSpctr2}.
  \item When $\proxy{x} =0$ and $d>0$, $\prox_{\alpha
  f}$  is multivalued in $0$ as all the points on the circle of radius
  $\rho^\best $ are solution of \Eq{eq:proxI-def}. As a consequence, $f$ is not
  \emph{prox-regular} at $\{0\}$ and its proximity operator is not defined for
  this point.
\end{itemize}

For practical reasons, we   define  $\prox_{\alpha
f}$ everywhere by assuming that $\angle(0) = 0$.
Thus, the proximity operator of $f$ %defined \Eq{eq:proxI-def}
is% $\prox_{\alpha f}(\proxy{x}) = \rho^\best \exp(\jmath\, \phi^\best)$  and
%  \begin{equation}
%  \phi^\best =  \left\{
%     \begin{array}{ll}
%       0 & \text{if } \proxy{x} = 0\,,\\
%       \proxy{\phi} & \text{otherwise}\,.\\
%     \end{array}\right.
%  \end{equation}
\begin{equation}
\prox_{\alpha\, f}(\proxy{x}) =  \left\{
\begin{array}{ll}
\rho^\best & \text{if } \proxy{x} = 0\,,\\
\rho^\best  \exp(\jmath\, \proxy{\phi})  & \text{otherwise}\,,\\
\end{array}\right.
\end{equation}
%where  $\rho^\best$ is given by the  positive  root of \Eq{eq:deriv}
%and $\proxy{\phi} = \angle \proxy{x}$.

Let us notice that the modified projection $P'(\proxy{x},d)$  defined by
Equation \ref{eq:MITA} lies also on the line   between $\proxy{x}$ and its
projection on the circle $\Abs{x}^2 = d$. Its position on this line depend on
the value of the relaxation parameter $\beta$. We can thus reinterpret this
modified projection as a heuristic approximation of the proximity operator.

\begin{figure*}[tb] \noindent\begin{minipage}[t]{.31\linewidth}\centering
\fbox{\includegraphics[height=\linewidth]{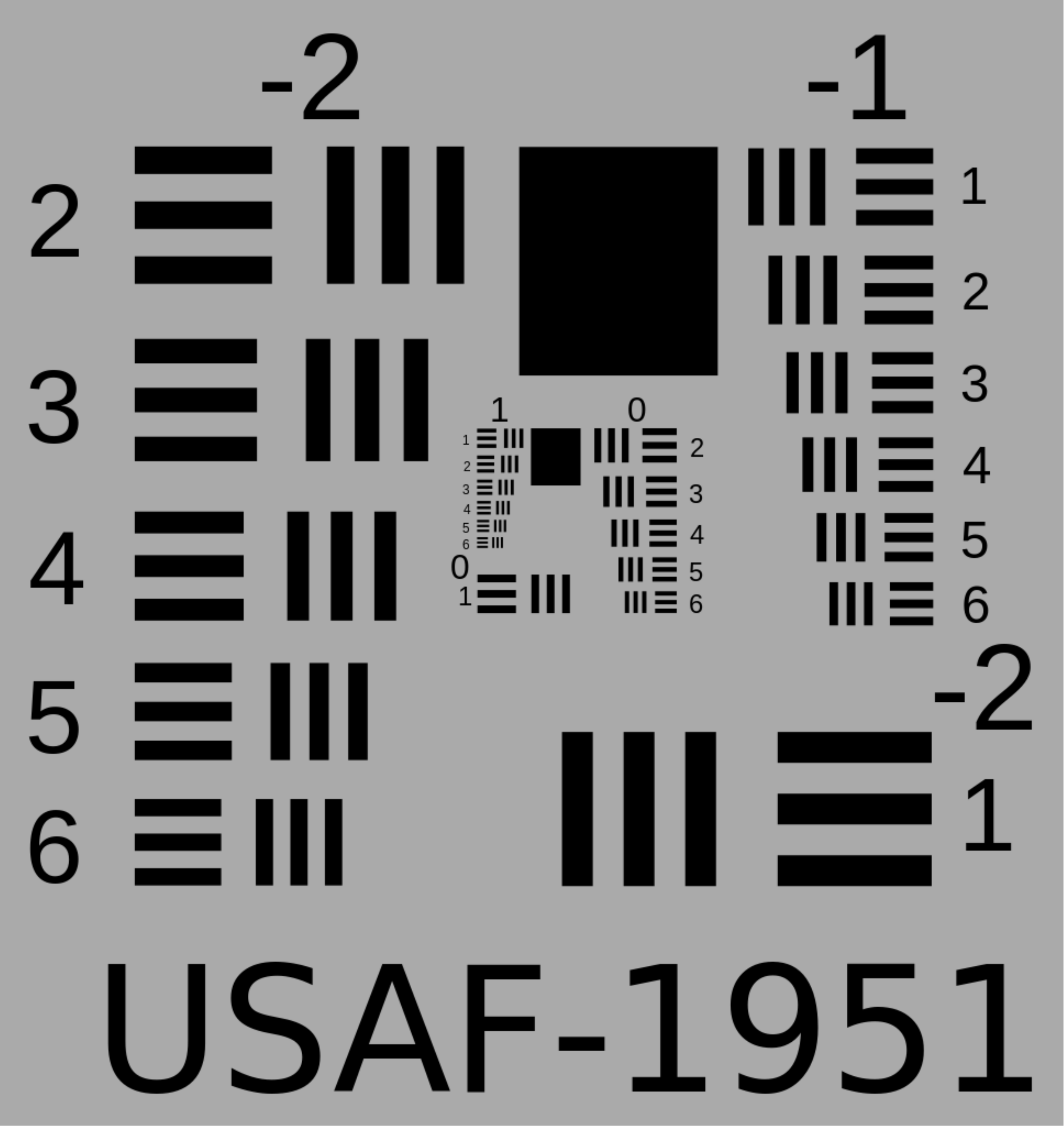}}
\caption{USAF-1951 test  image used.}
\label{fig:USAF}
\end{minipage}\hfill
\begin{minipage}[t]{.31\linewidth}\centering
\fbox{\includegraphics[height=\linewidth]{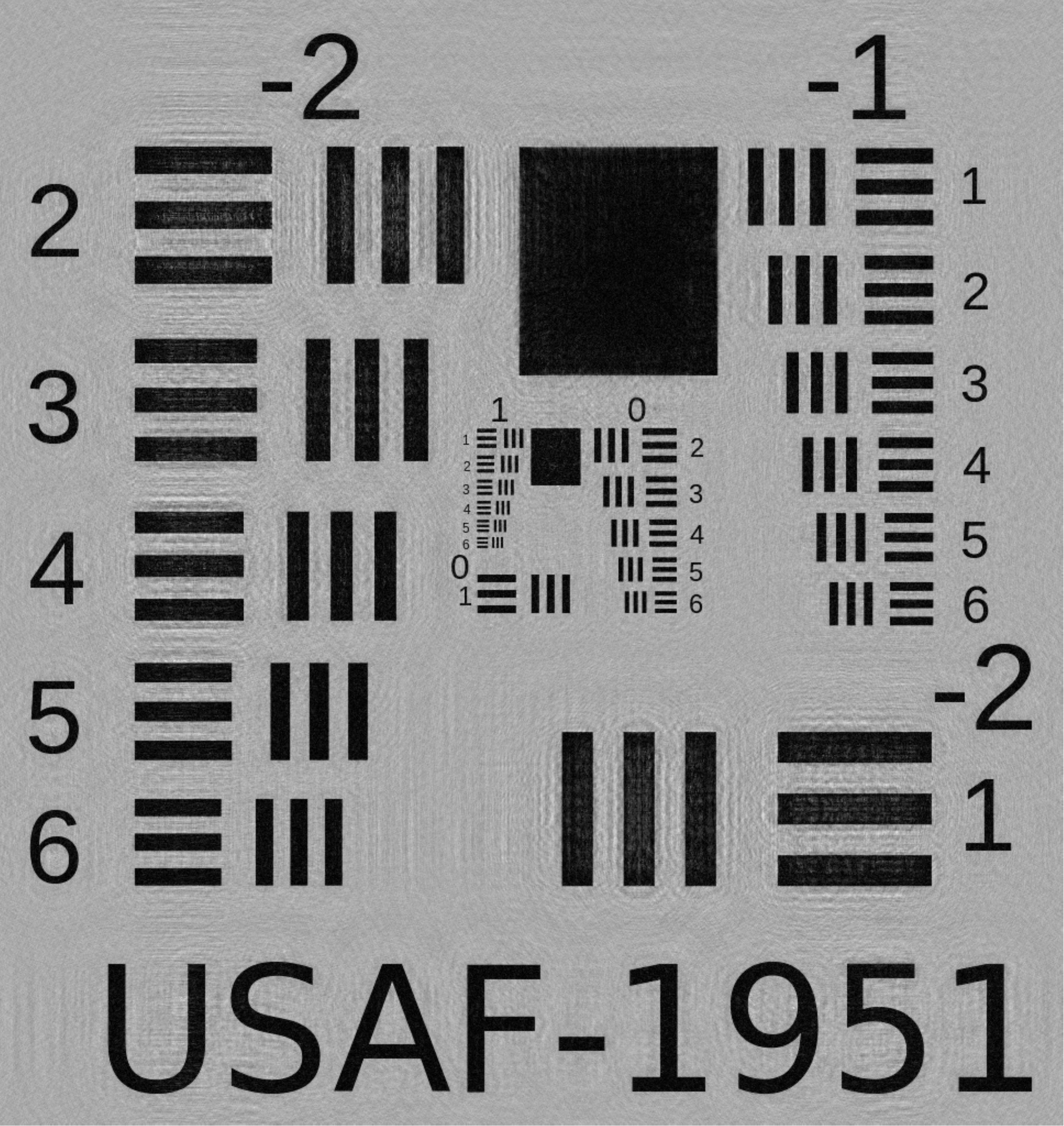}}
\caption{Intensity of the estimated wave using the
classical projection (noiseless case) back-propagated from $z_A$ to $z_0$.
 $\SNRxb = 15.12\,$dB.} 
\label{fig:CPs0}
\end{minipage}\hfill
\begin{minipage}[t]{.31\linewidth}\centering
\fbox{\includegraphics[height=\linewidth]{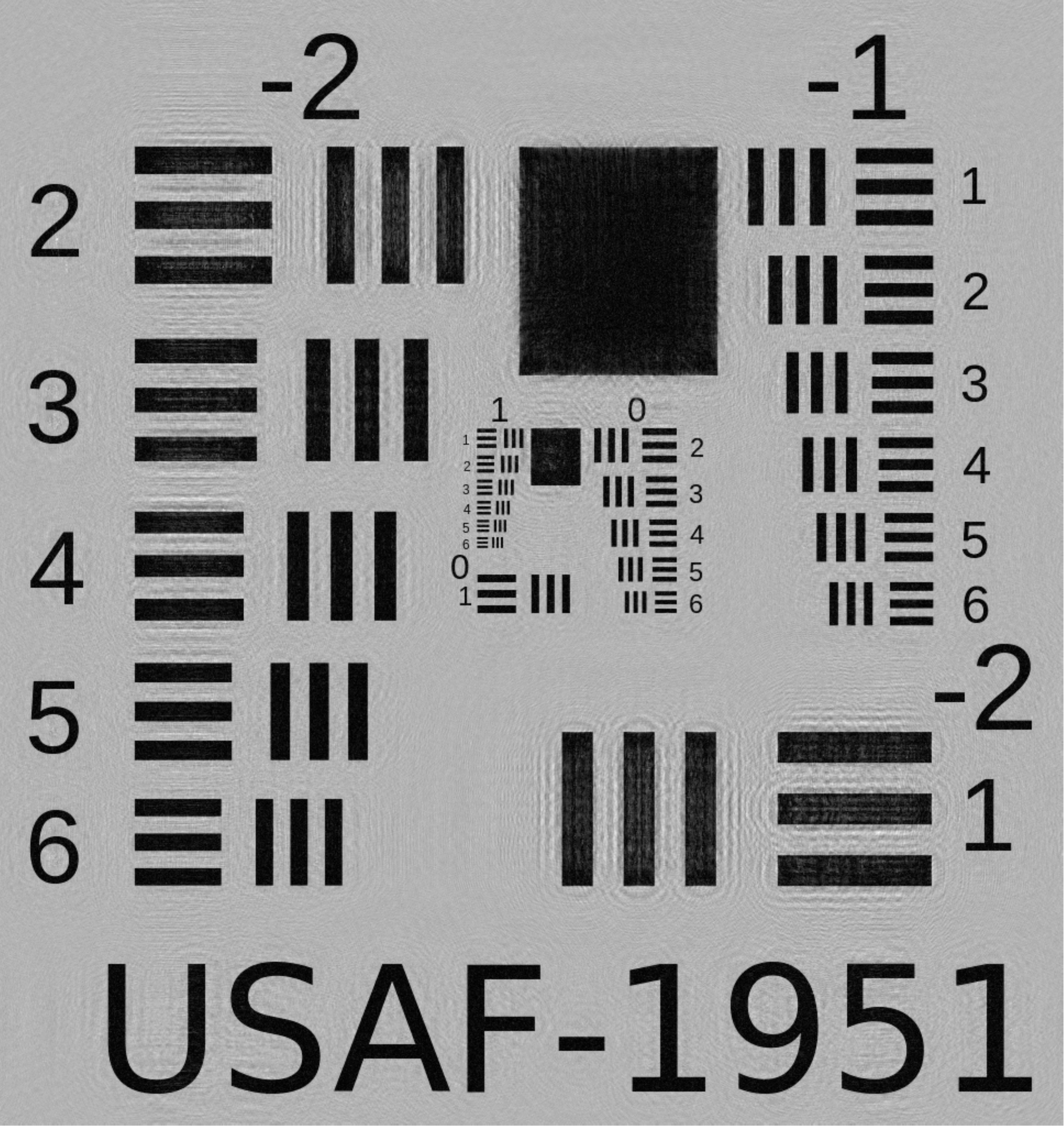}}
\caption{Intensity of the estimated wave using the proposed proximity operator
(noiseless case)  back-propagated from $z_A$ to $z_0$.
$\SNRxb = 15.68\,$dB.}
\label{fig:GPs0}
\end{minipage}
\begin{minipage}[t]{.31\linewidth}\centering
\fbox{\includegraphics[height=\linewidth]{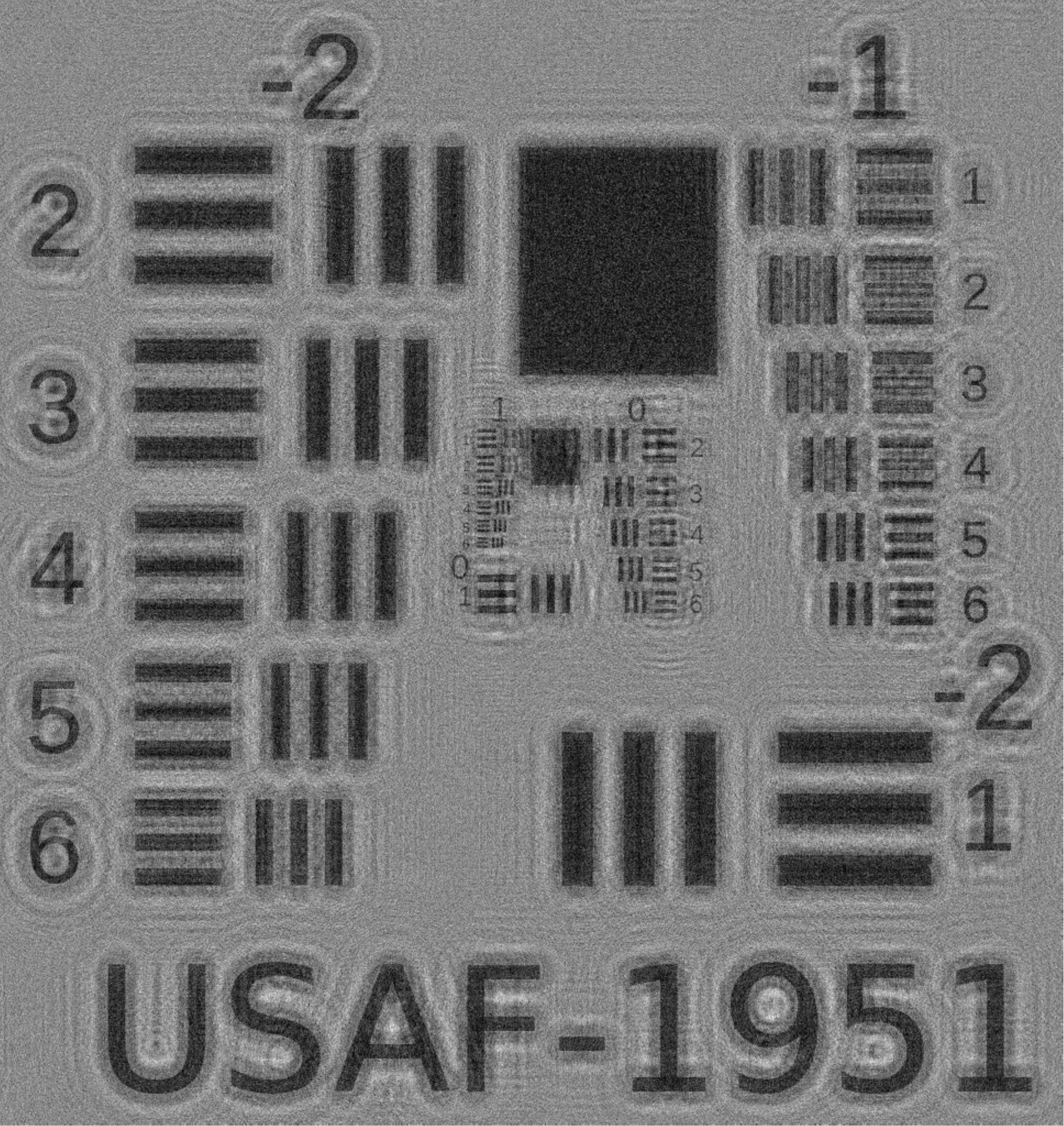}}
\caption{Intensity of the estimated wave using the
classical projection (noise standard deviation $\sigma=0.3$) back-propagated
from $z_A$ to $z_0$.
$\SNRxb = 6.37\,$dB.}
\label{fig:CPs3}
\end{minipage}\hspace{.02\linewidth}
\begin{minipage}[t]{.31\linewidth}\centering
\fbox{\includegraphics[height=\linewidth]{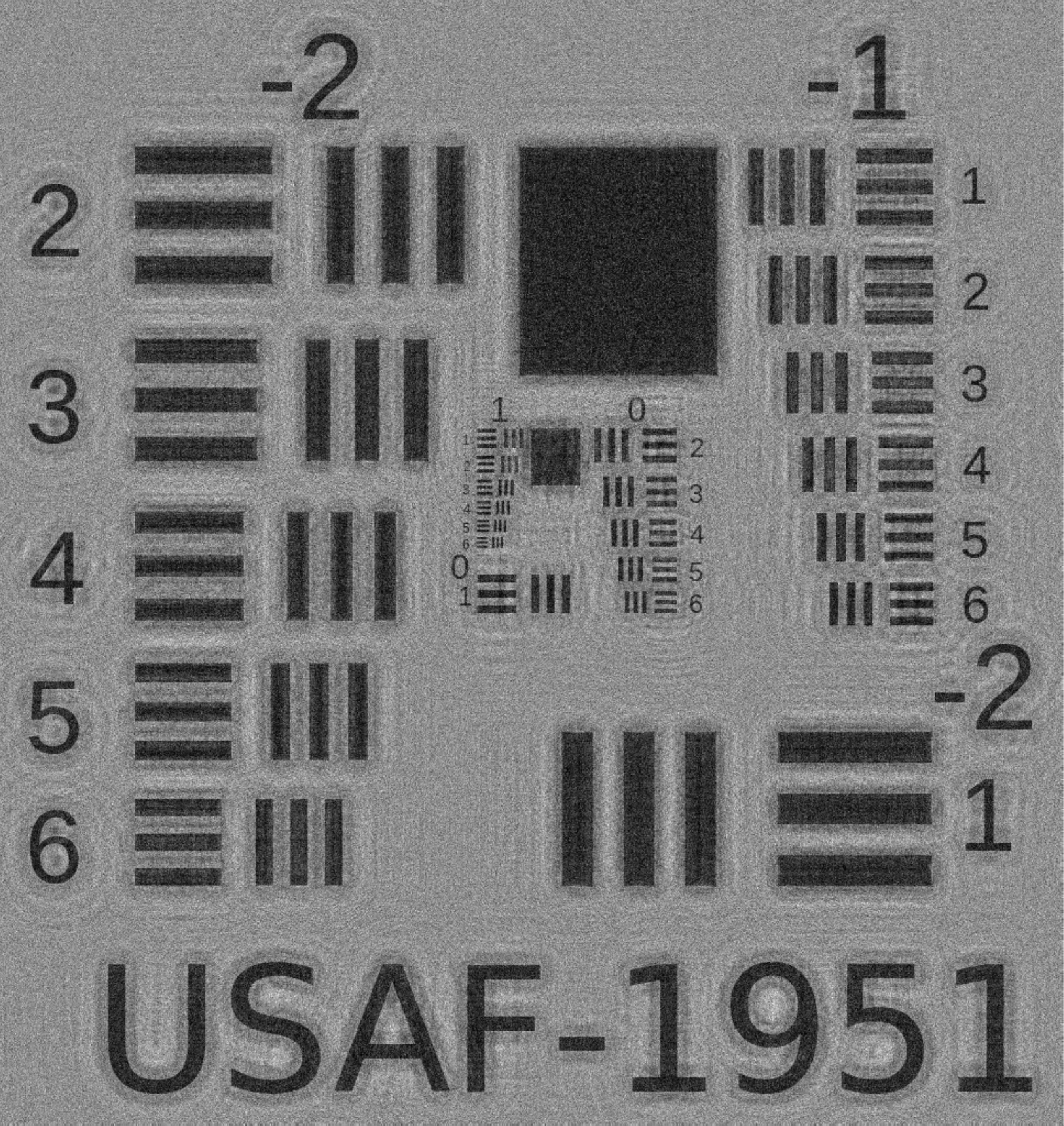}}
\caption{Intensity of the estimated wave using the proposed proximity operator
(noise standard deviation  $\sigma=0.3$)   back-propagated from $z_A$ to $z_0$.
$\SNRxb = 7.60\,$dB.}
\label{fig:GPs3}
\end{minipage} \hspace{.02\linewidth}
\begin{minipage}[t]{.33\linewidth}
\fbox{\includegraphics[height=0.72\linewidth]{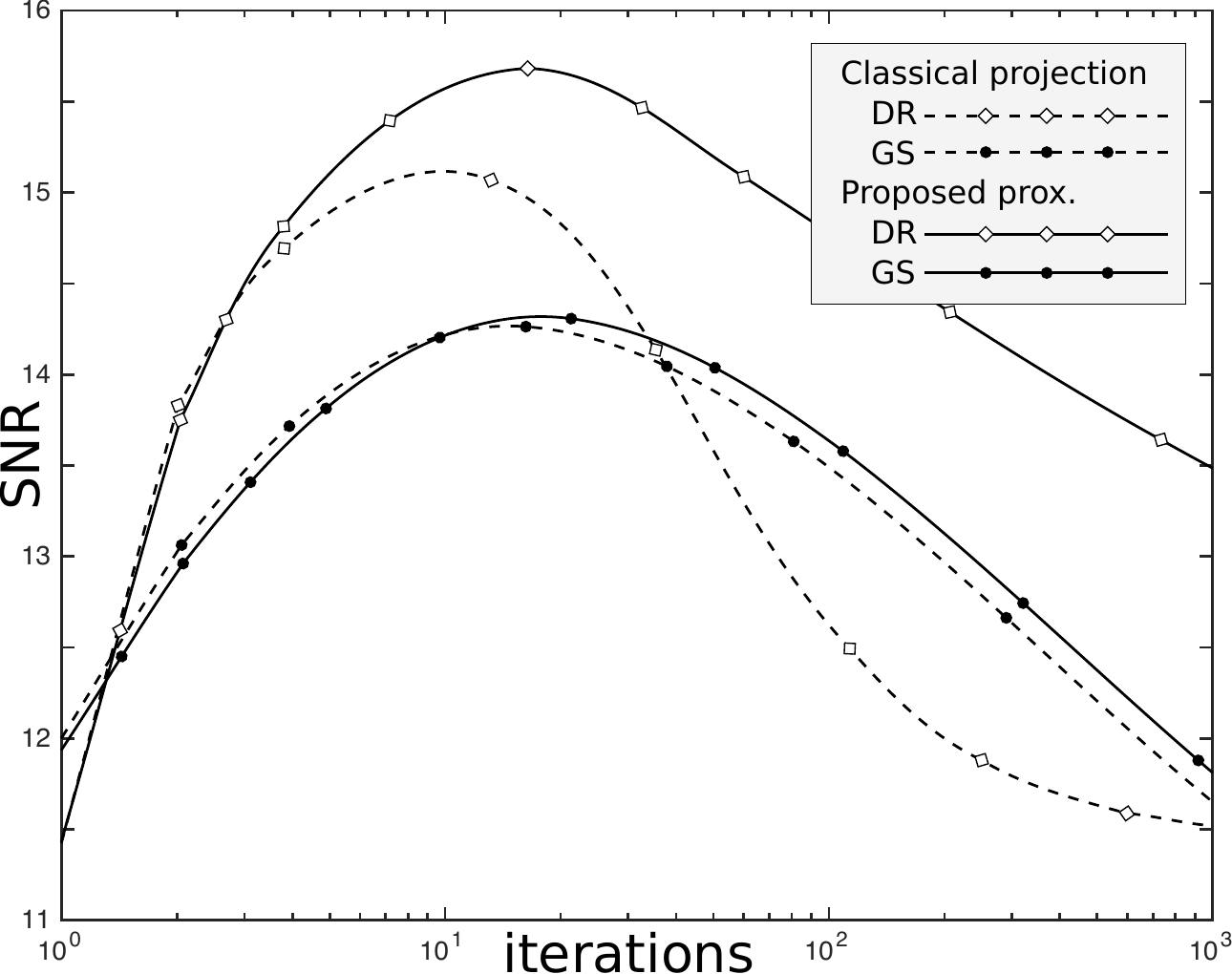}}
\caption{Comparison of DR and GS performance without noise using the classical
projection or the proposed operator}
\label{fig:RMSDG0}
\end{minipage}
\end{figure*}

\begin{figure*}[tb]
\noindent\begin{minipage}[t]{.33\linewidth}
\fbox{\includegraphics[height=.72\linewidth]{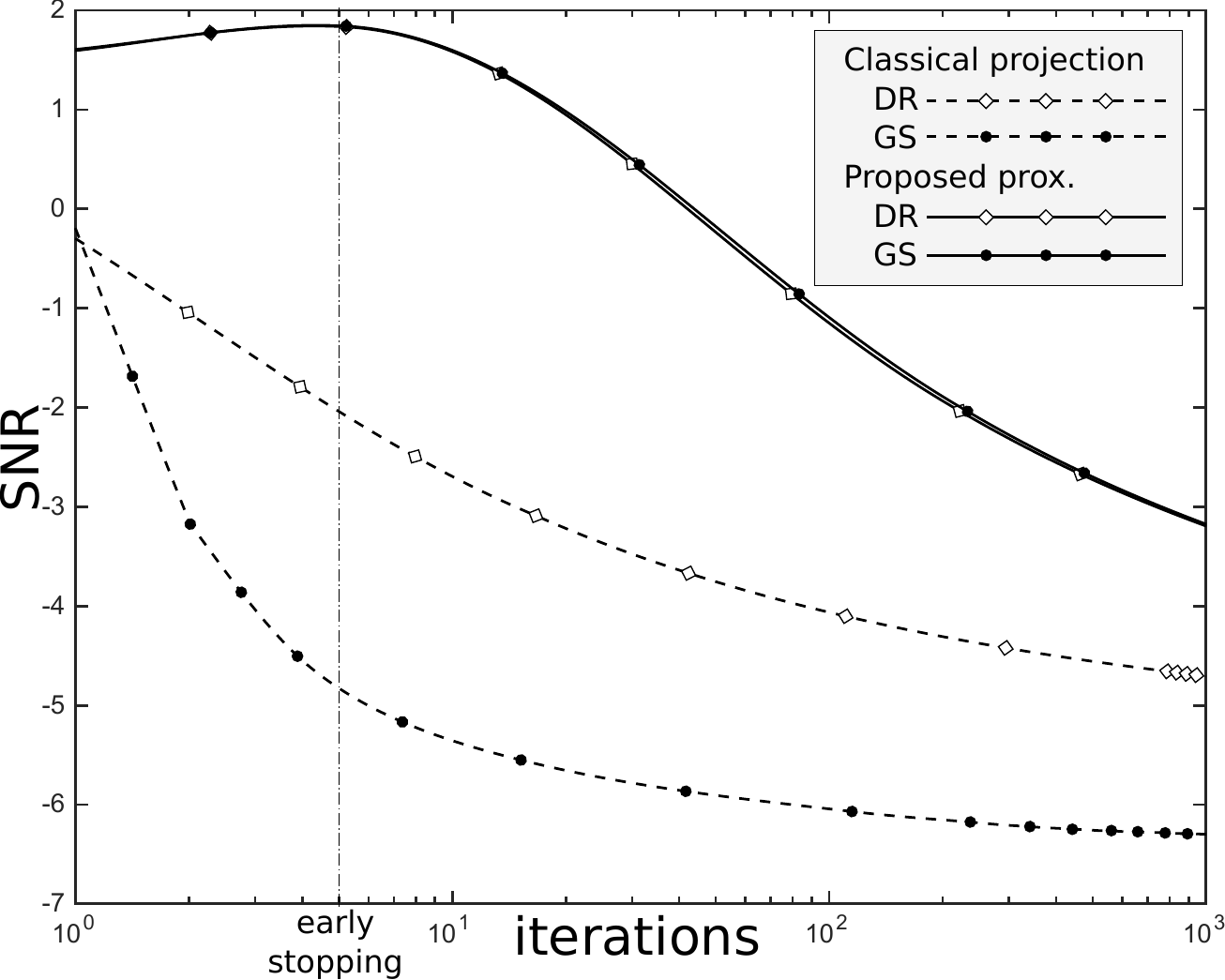}}
\caption{Comparison of DR and GS performance in noisy conditions ($\sigma=1$)
using the classical projection or the proposed operator.}
\label{fig:RMSDG1}
\end{minipage}\hfill
\begin{minipage}[t]{.33\linewidth}
\fbox{\includegraphics[height=.72\linewidth]{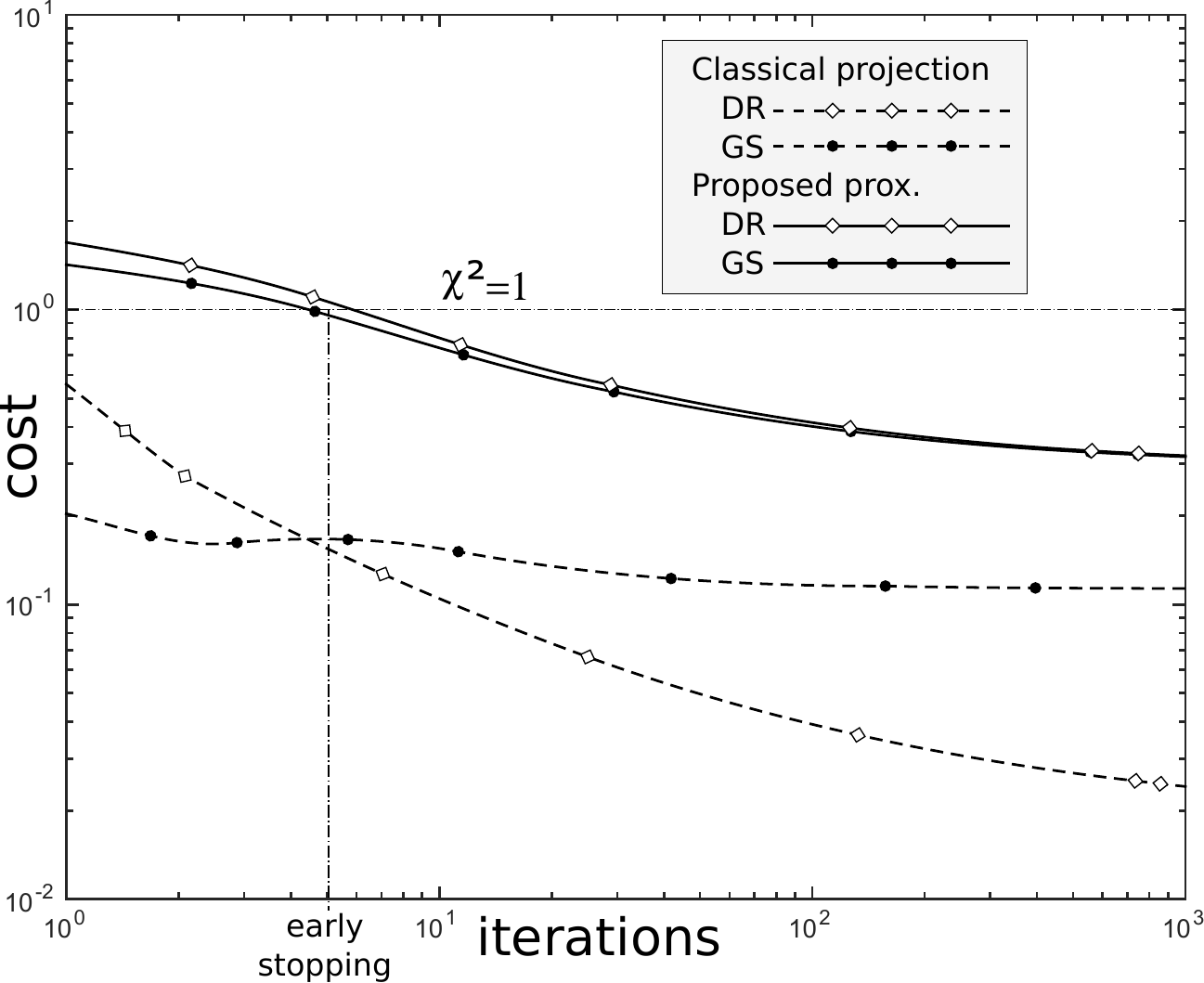}}
\caption{Evolution of the cost function for both algorithms and both
projectors.}
\label{fig:costDG0}
\end{minipage}\hfill
\begin{minipage}[t]{.31\linewidth}\centering
\fbox{\includegraphics[height=0.78\linewidth]{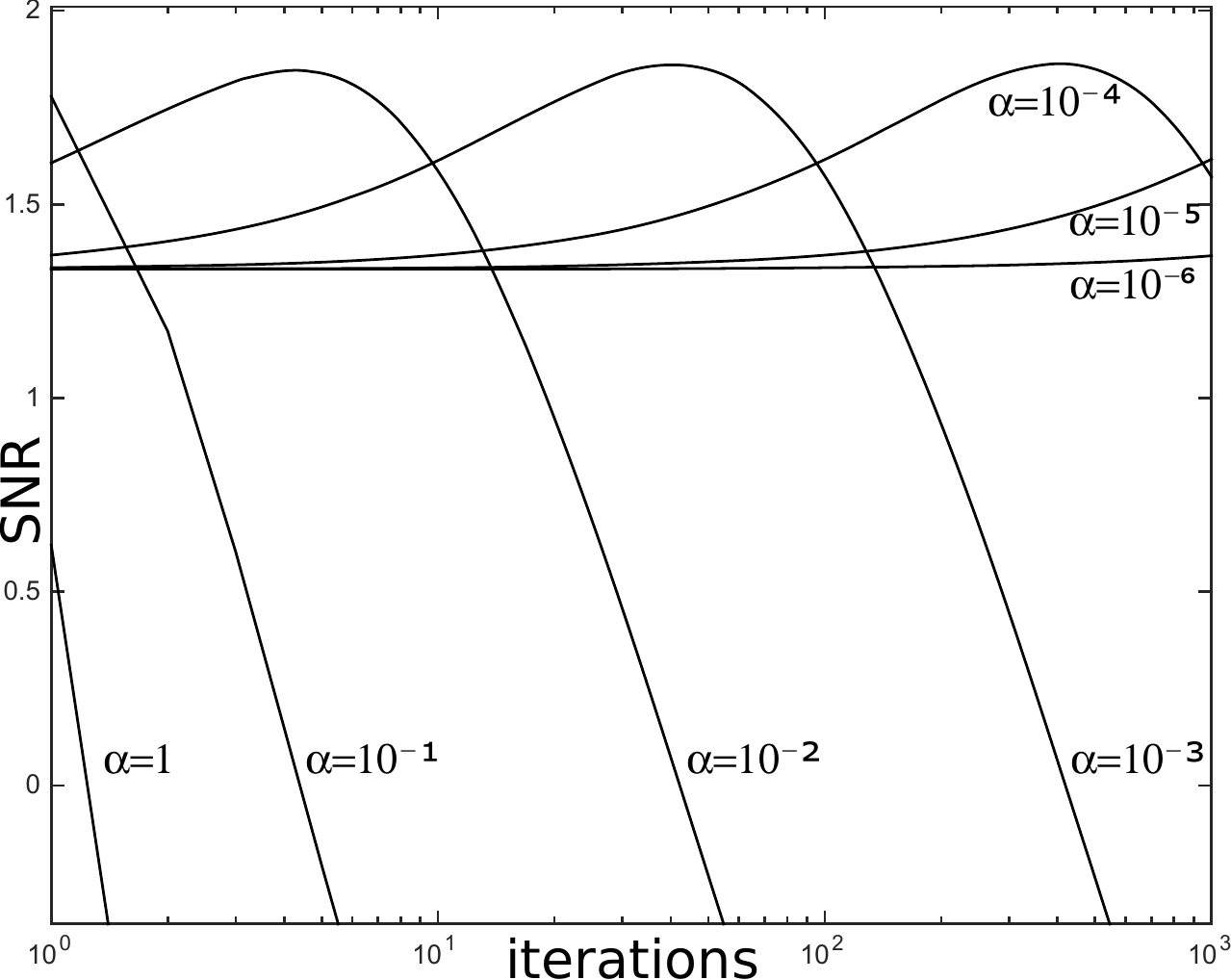}}
\caption{Performance of DR ($\sigma=1$) with the proposed proximity operator
$\prox_{\alpha f}$ for different values of $\alpha$.}
\label{fig:Alpha}
\end{minipage}
\end{figure*}

\subsection{Gaussian Likelihood}
\label{sec:Gauss}
For additive  Gaussian noise at a given pixel with variance $\sigma =
\Var\{{d}\}$, the function $f$ writes
\begin{equation}
f(x) = w (\Abs{x}^2 -
d)^2\,,\label{eq:Glkldef}
\end{equation}
where $w = 1/\sigma^2$ is the inverse variance of the noise at the considered
pixel.
In this case,  \Eq{eq:rho} becomes:
\begin{equation}
\rho^\best = \argmin_{\rho\ge 0} \, \left(  \alpha\,w \,(\rho^2 -d)^2 + \frac{1}{2}\,(\rho -\proxy{\rho})^2 \right)\, .
\end{equation}
The solution  is then one of the  roots
of the polynomial $q_G$ defined as
\begin{eqnarray}
q_G(\rho) &=& \frac{\D{} }{\D{ \rho}} \left( \alpha\,w \,(\rho^2 -d)^2  +
\frac{1}{2}\,(\rho -\proxy{\rho})^2\right) \nonumber\\
 &=& 4\,\alpha\,w\,\rho^3 + \rho \left(1 - 4\,\alpha\,wd\right)
- \proxy{\rho} \,.\label{eq:deriv}
\end{eqnarray}
As there is no second coefficient in this cubic polynomial, 
the sum of its roots is zero whereas their product is strictly positive since
${\proxy{\rho}}/( 4\,\alpha\,w)>0 $. Thus, $q_G$ has always only one
positive root $\rho^+$.
As stated in the previous section, this root must
 lie between $\sqrt{d}$ and $\proxy{\rho}$. It is 
computed using Cardano's  method.

% In the (physically impossible) case where \Eq{eq:deriv}
%has no positive roots, then the polynomial is strictly increasing for $\rho>0$
% and $\rho^\best = 0$.

\subsection{Poisson Likelihood}
\label{sec:poisson}
In the photon-counting case, the noise follows a Poisson distribution and the
function $f$ writes
\begin{equation}
f(x) =  \Abs{x}^2 - d\,\log\left( \Abs{x}^2 + b \right)
\,,\label{eq:Plkldef}
\end{equation}
where $b$ is the expectation of some spurious independent Poisson process that
accounts for background emission and detector dark current at the considered
pixel.
Given this noise distribution, the solution of \Eq{eq:rho} is given by the largest
real root of the cubic polynomial $q_P(\rho) =\frac{\D f(\rho)}{\D \rho}$, with
\begin{align}
q_P(\rho)&=\frac{\D{} }{\D{ \rho}} \left( \alpha\,f(\rho) +
\frac{1}{2}\,(\rho -\proxy{\rho})^2\right) \nonumber\\
  &= (2\,\alpha+1)\,\rho^3 -  \proxy{\rho} \,\rho^2 + \left((2\,\alpha
+ 1)\,b - 2\,\alpha\,d\right)\,\rho -b\,\proxy{\rho} \,.\label{eq:ph3rd}
\end{align}
As in the case of \Eq{eq:deriv}, this root is computed using Cardano's method. 
When no background emission is present ($b=0$), this polynomial reduces to a
quadratic equation whose largest root always exists and is given by
\begin{equation}\label{eq:ph2nd}
\rho^\best =  \frac{\proxy{\rho}+ \sqrt{8\,d\,\alpha\,(1 +
2\,\alpha)+\proxy{\rho}^2}}{2 + 4\, \alpha}\,.
\end{equation}

\section{Proximity operator for a sum of intensity measurements}
\label{sec:SumProx}
In this section, we extend the presented proximity operators 
to the case where $N$  complex amplitudes sum up incoherently on a 
pixels. This corresponds to the multispectral case or when interference fringes
exhibit high frequencies that are  not sufficiently sampled by the
detector. In this case, an appropriate  forward model is
\begin{equation}
d_k = \Norm{\V{y}_{k}}_2^2 + n_k\,,
\end{equation}
where $\V{y}_k \in \Complexes^{N}$  is a vector containing the $N$
complex amplitudes arriving on the pixels $k$.  In the undersampled-fringes
case,  this vector writes $\V{y}_k  = (x_{N(k-1)+1},\dots,x_{Nk})$, where the
factor $N$ is chosen  such that the
adequately sampled complex amplitude $\V{x} \in \Complexes^{NK}$  fulfills the Nyquist criterion.
With this forward model, the
likelihood function writes $\ell_k(\Norm{\V{y}_k}^2 \,;\, d_{k})$. By setting
$\V{y}_k = \eta\,\V{u}$,  with $\eta\ge0$ and $\Norm{\V{u}}_2=1$, we can define $\alpha\,f(\eta) = 
\ell_k(\eta^2 \,;\, d_{k})$.
Then \Eq{eq:proxI-def} becomes
\begin{equation}
\prox_{\alpha f}(\proxy{\V{y}})  = \argmin_{\eta\ge0\ ,\ \Norm{\V{u}}_2=1}
\left( \alpha\, f\left( \eta \right) + \frac{1}{2} \,
\Norm{\eta\,\V{u} - \proxy{\V{y}}}^2_2 \right)  \, .
\label{eq:Undersp}
\end{equation}
% 
% 
% \begin{equation}
% \prox_{\alpha f}(\V{y})  = \argmin_{\V{x}}  \left\{\alpha\,
% f\left( \Norm{\V{x}}_2^2 - I\right) + \frac{1}{2} \,
% \Norm{\V{x} - \V{y}}^2_2 \right\}  \, .
% \label{eq:Undersp}
% \end{equation}
% By setting $\V{x} = \eta\,\V{u}$  with $\eta\ge0$ and
% $\Norm{\V{u}}_2=1$, this can be written:
% \begin{equation}
% \prox_{\alpha f}(\V{y})  = \argmin_{\eta\ge0\ ,\ \Norm{\V{u}}_2=1} \left\{
% \alpha\, f\left( \eta^2 - I\right) + \frac{1}{2} \,
% \Norm{\eta\,\V{u} - \V{y}}^2_2 \right\}  \, .
% \label{eq:Undersp}
% \end{equation}
%
% \begin{equation}
%    \min_{\V{u},\eta} \alpha\, \left( \eta^2 -
%   I\right)^2 + \frac{1}{2} \, \Norm{\eta\, \V{u} - \V{y}}^2_2
%   \quad\text{s.t.}\quad
%   \eta \ge 0
%   \quad \text{and}\quad
%     \Norm{\V{u}} = 1\, .
%   \label{eq:prob2w}
% \end{equation}
%
By assuming that  $\Norm{\proxy{\V{y}}}_2 \not= 0$ and $\eta>0$, we find that
\begin{equation}
\V{u}^\best(\eta)
= \argmin_{\V{u}, \Norm{\V{u}} = 1}
\Norm{\eta\,\V{u} - \proxy{\V{y}}}^2_2 = \frac{\proxy{\V{y}}}{\Norm{\proxy{\V{y}}}_2} \, .
\label{eq:best-u}
\end{equation}
Thus the solution is
\begin{equation}
\V{y}^\best = \eta^\best \, \frac{\proxy{\V{y}}}{\Norm{\proxy{\V{y}}}_2} \, ,
\label{eq:solution}
\end{equation}
where   $\eta^\best$ is given by
\begin{eqnarray}
\eta^\best &=& \argmin_\eta  \left( \min_{\V{u}, \Norm{\V{u}} = 1} \left(
\alpha\, f\left( \eta \right) + \frac{1}{2} \,
\Norm{\eta\,\V{u} - \proxy{\V{y}}}^2_2 \right) \right)  \, ,\\
&=& \argmin_{\eta>0}\left(   \alpha\,f\left(\eta\right)
+ \frac{1}{2} \, \left({\eta} - \Norm{\proxy{\V{y}}}_2\right)^2 \right)\, ,
\label{eq:EtaSubSamp}
\end{eqnarray}
since \begin{equation}
\min_{\V{u}, \Norm{\V{u}} = 1}
\Norm{\eta\,\V{u} - \proxy{\V{y}}}^2_2
= \left(\Abs{\eta} - \Norm{\proxy{\V{y}}}_2\right)^2 \, .
\label{eq:min_2nd-term}
\end{equation}
Solving \Eq{eq:EtaSubSamp} is equivalent to solving \Eq{eq:rho} with
$\proxy{\rho} = \Norm{\proxy{\V{y}}}_2 $. In the case where $\Norm{\proxy{\V{y}}}_2 = 0$ and
$d>0$, $f$ is not prox-regular and \Eq{eq:Undersp} has an infinite number of
solutions.  As in Section~\ref{sec:proxlkl}, we assume in practice that
$\prox_{\alpha\,f}{(\proxy{\V{y}})} = \eta^+$ when $\Norm{\proxy{\V{y}}}_2 = 0$.
 To sum up, the proximity operator for undersampled measurements is:
\begin{equation}
% \rho^\best &=& \argmin_{\rho\ge 0} \alpha\,  w
% (\rho^2 - I)^2 + \frac{1}{2}\,(\rho -
%   \proxy{\rho})^2 \, ,\label{eq:rho}\\
\prox_{\alpha\,f}{(\proxy{\V{y}})} =  \left\{
\begin{array}{ll} \eta^\best\,, & \text{if } {\Norm{\proxy{\V{y}}}_2} = 0\,\\
 \eta^\best  \frac{\proxy{\V{y}}}{\Norm{\proxy{\V{y}}}_2}\,, &
 \text{otherwise}\,.\\
 \end{array}\right.
% 
% \frac{\proxy{\V{y}}}{\Norm{\V{y}}_2} \,\argmin_{\eta\ge 0} \, \left\{
% \alpha\,f (\eta) + \frac{1}{2}\,(\rho -
% \proxy{\rho})^2 \right\}\, ,\label{eq:undr}\\
% \V{u} &=&  \proxy{\phi}\,,
\end{equation}
%where $ \proxy{x} =   \proxy{\rho} \exp(\jmath\, \proxy{\phi})$.
This proximity operator for undersampled intensity
measurements can be computed for any  function  $f$ that  has a  proximity
operator in closed form such the Gaussian or Poisson likelihood described in the
previous sections.

\section{Numerical experiments}

To study the performance of the proposed proximity operators, we 
simulated  one of the simplest setup of phase retrieval. Under a Fresnel
approximation, we simulated numerically  a  wave diffracted by a planar real
object (here a $K = 1024\times984$ pixels image of the USAF resolution test
chart shown \Fig{fig:USAF}) placed at $z_0 = 0$. The diffracted wave at $z_A$ is
the reference complex amplitude $\V{r}$ that will be estimated throughout the experiments.
We computed the noisy intensities  $\V{d}_A = \Abs{\V{r}} + \V{n}_A$ and
$\V{d}_B = \Abs{\M{H}\,\V{r}} + \V{n}_B$ at  depth $z_A$ and $z_B$, where
$\M{H}$ is the propagation operator from $z_A$ to $z_B$ and $\V{n}_A$ and
$\V{n}_B$ are  noise vectors with identical statistics given by the
experimental conditions.
The setup parameters are:
$\lambda= 633\,$nm, pixel size:
$=5.3\,$\micron, $z_A = 1\,$cm, and $z_B = 2\,$cm.

For each experiment, we built the functions $f_{A,k}(x)=\ell_k(\Abs{x}^2 \,;\,
d_{A,k})$ and  $f_{B,k}(x)=\ell_k(\Abs{x}^2 \,;\,
d_{B,k})$ according to the considered noise model.
 We then compared the performance of the proposed proximity operator
 $\prox_{\alpha f}$ to that of the classical projection defined  by \Eq{eq:StrictConstraint}   by estimating
the complex amplitude of the wave $\V{x}^\best$ at $z_A$. To keep the problem as
simple as possible, we only used the knowledge of measured intensities
 without additional prior (neither regularization, nor use of the fact
that the image is non-negative at $z_0$).

 In all experiments, the quality of
the recovered complex amplitude $\V{x}$ in plane $z_A$ is assessed by the mean
of the reconstruction signal to noise ratio:
\begin{equation}
\SNR{\V{x}} = 10\,\log_{10}  \frac{\Norm{\V{r}}^2_2}{\Norm{\V{r} -
\V{x}}^2_2}\,.
\end{equation} 
 As the
initial wave is real in the plane $z_0=0$, back-propagating the estimated wave
from $z_A$ to $z_0$ is used as a visual assessment of the reconstruction quality
as shown Figures \ref{fig:CPs0} to \ref{fig:GPs3}.
Let us remind that as the phase retrieval problem is not convex, the solution
depends on the initialization. We chose the initialization $\V{x}^{(0)} =
\sqrt{\V{d}_A}$ for every experiments and a different initialization may lead to
a different recovered complex amplitude with a different SNR.

\subsection{Alternating Projection or Douglas Rachford?}

\begin{algorithm}
\caption{Douglas-Rachford algorithm}\label{alg:DR}
\begin{algorithmic}[1]
\Procedure{DR}{$f_A$, $f_B$}
\State $\V{y}^{(0)} = \sqrt{\V{d}_A}$ and $\lambda\in ]0,2[$
\Comment{init. ($\lambda=1$ for all results)}
\For{$n =  1, \dots, \textrm{maxiter}$}
\State $\V{x}^{(n)} = \prox_{\alpha f_A}(\V{y}^{(n-1)})  $ %\Comment{Propagation
% to the $z_B$ plane}
\State $\V{r}^{(n)} = 2\,\V{x}^{(n)} - \V{y}^{(n-1)}  $
\State $\V{y}^{(n)} = \V{y}^{(n-1)} +  \lambda \left( \M{H}^{\T} 
\prox_{\alpha f_B} \left( \M{H}\, \V{r}^{(n)}
\right) - \V{x}^{(n)} \right) $
%\Comment{Projection}
\EndFor\label{GSendwhile}
\State \textbf{return} $\V{x}^{( \textrm{maxiter})}$\Comment{Complex
amplitude in the $z_A$ plane}
\EndProcedure
\end{algorithmic}
\end{algorithm}

\begin{figure*}[htbp] 
\noindent\begin{minipage}[t]{.31\linewidth}\centering
\fbox{\includegraphics[height=.78\linewidth]{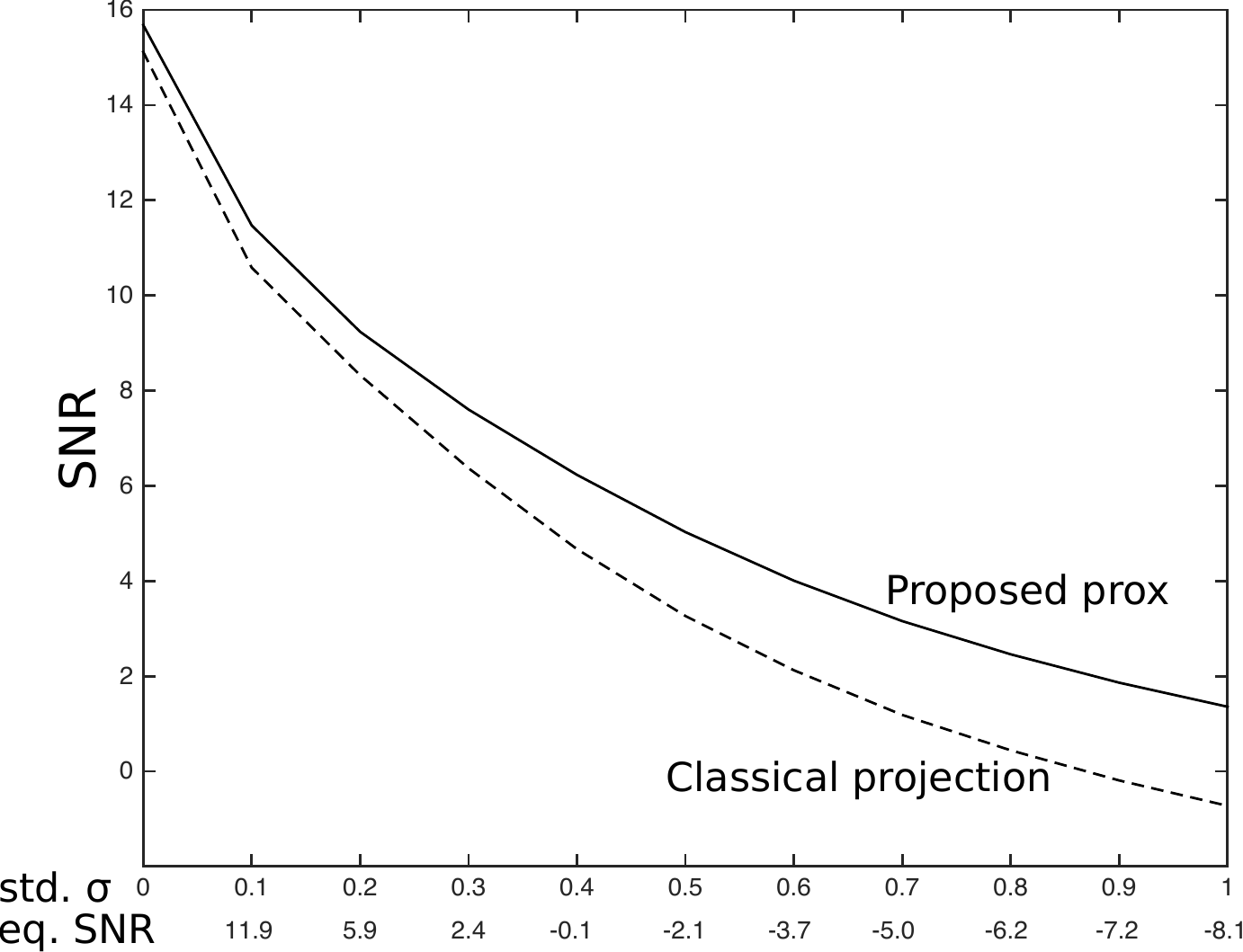}}
\caption{Comparison of both projectors using the DR algorithm as a function of
noise (noise level given in standard deviation and SNR).}
\label{fig:Gaussian}
\end{minipage}\hfill
\begin{minipage}[t]{.31\linewidth}\centering
\fbox{\includegraphics[height=.78\linewidth]{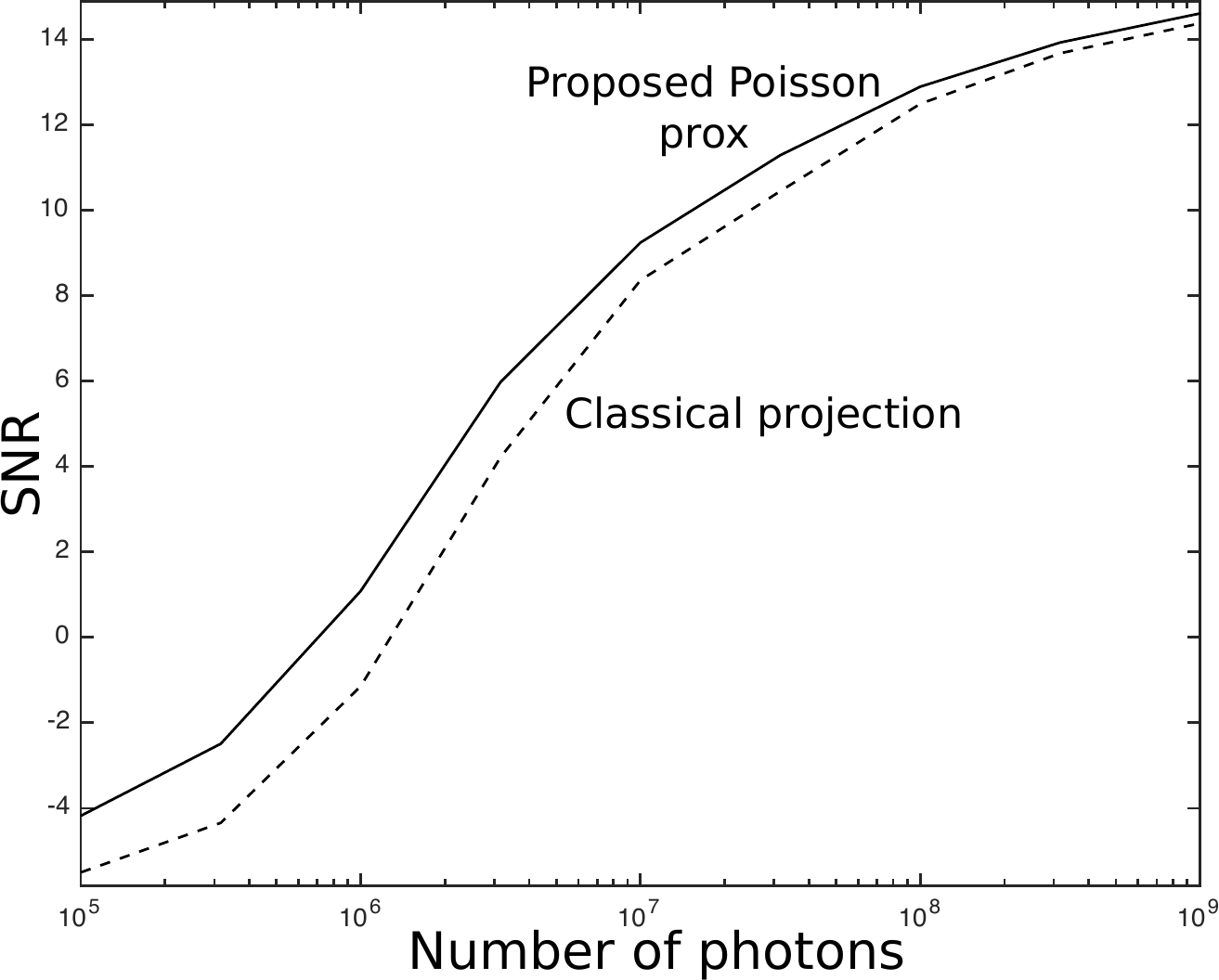}}
\caption{Comparison of classical projection and Poisson proximity as a function
of the number of  photons ($10^5$ photons  $\approx 1$ photon per pixel, on
average).}
\label{fig:Photon}
\end{minipage}\hfill
\begin{minipage}[t]{.31\linewidth}\centering
\fbox{\includegraphics[height=0.8\linewidth]{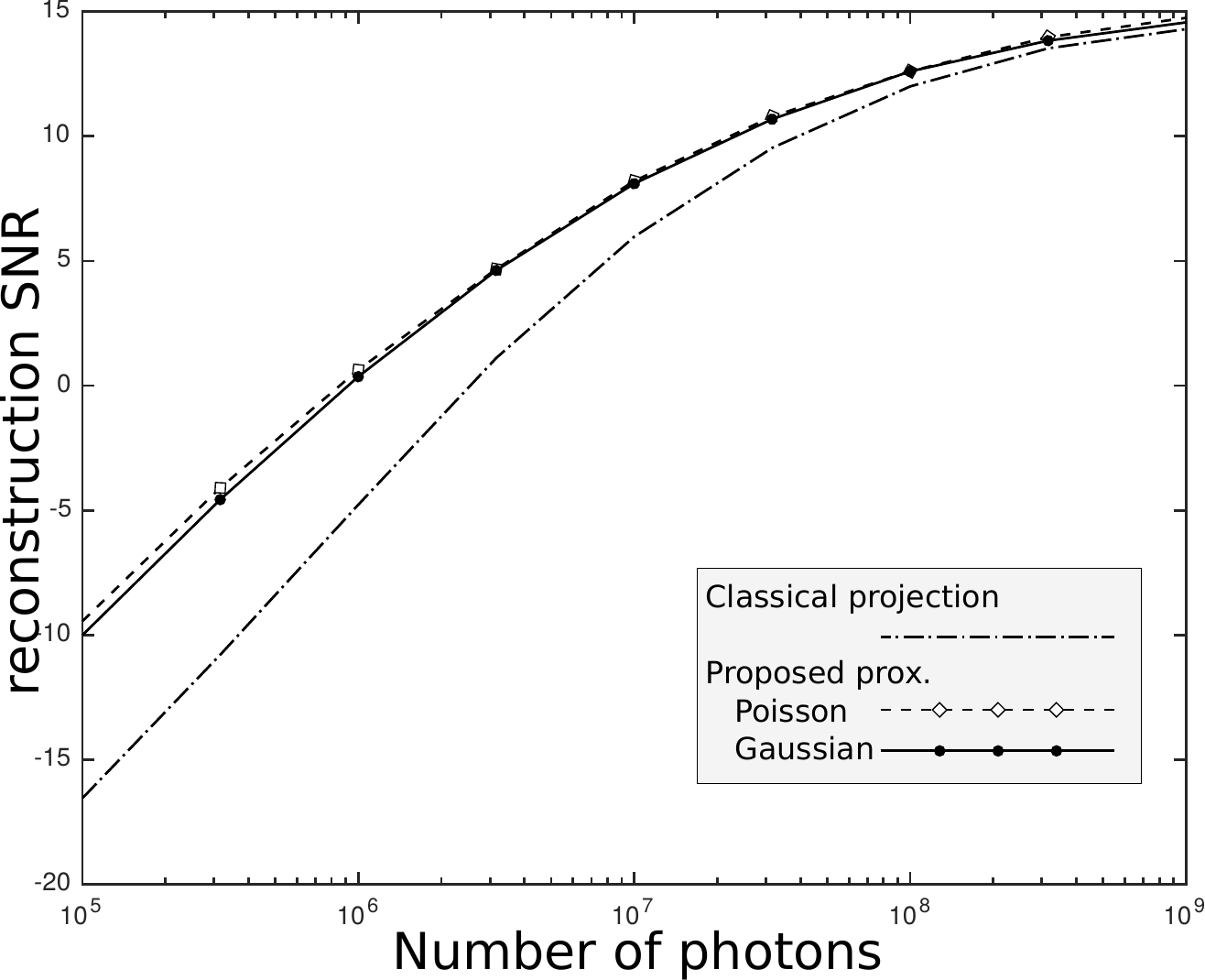}}
\caption{$\SNRxb$ as  function of the number of photons in presence
of a $3\,e^-$ dark current.}
\label{fig:LLRMS}
\end{minipage}
\end{figure*}

The use of  the proposed operator in   Algorithm \ref{alg:GS} instead
of the classical projection $P_A$ and $P_B$ amounts to solving
\begin{eqnarray}
\V{x}^\best &\in&\argmin_{\V{x} \in \Complexes^{K}} \left(
\sum^N_{k=1} f_{A,k}\left(x_k\right)\right. \nonumber \\ 
&& + \left. \inf_{\V{y} \in \Complexes^{K}}\left(\sum^K_{k'=1}
f_{B,k}\left(y_{k}\right) +
\frac{1}{2} \Norm{\M{H}\,\V{x} - \V{y}}_2^2\right)\right)\,,\label{eq:MAP}
\end{eqnarray}
which is a relaxed version of \Eq{eq:I-lkl}. Alternatively,
\Eq{eq:I-lkl} can be solved using the Douglas-Rachford (DR)
algorithm described in Alg.~\ref{alg:DR} thanks to the following property on
the proximity operator of $g(\V{x}) = f(\M{H}\cdot\V{x}) $  \cite{CombettesPesquet2011}:
\begin{equation}
\M{H}\cdot\M{H}^{\T} = \M{Id}
\Longrightarrow  \prox_{\alpha g}(\V{x}) = \M{H}^{\T}\cdot \prox_{\alpha f} (
\M{H}\cdot \V{x})\,,
\end{equation}
where $\M{Id}$ is the identity matrix.

For the Gaussian likelihood as for the Poisson likelihood,  $f$ is not convex.
The convergence of  both algorithms cannot be proved  even if there exist some
convergence results in the related case of the estimation of the intersection of
a circle and a line \cite{HesseLuke2013}. The solution may therefore depend on
the starting point. In all the presented experiments, we begin with the starting amplitude in
$z_A$ plane $x^{(0)}_A = \sqrt{d_A}$.

With the classical projection, % \Eq{eq:StrictConstraint},
DR is more efficient
than GS as  can be seen in \Fig{fig:RMSDG0} and \Fig{fig:RMSDG1}, either with
or without noise. In the presence of noise and using the proposed proximity
operator, the performances of both algorithms are  similar; they  become
indistinguishable as the amounts of noise level increases.

\subsection{Tuning the Parameters}

With the proposed proximity operator, two parameters have to
be tuned: the number of iterations and the parameter
$\alpha$. All tests with the DR algorithm were done with $\lambda=1$.

Phase retrieval is an ill-posed problem. The number of unknowns ($2\,K$)
is equal to the number of measurements,  meaning that  such
\emph{maximum-likelihood} algorithms are subject to noise amplification. Hence, 
$\SNRx$ began to worsen 
after some iteration, while the cost was still decreasing, 
as  can be seen \Fig{fig:RMSDG1} and \Fig{fig:costDG0}.
The correct prescription of the number of iterations is essential to stop the
algorithm at the precise moment when the wavefront gives the best SNR.  This 
 is classically done in phase retrieval and acts as a regularization
\cite{Luke2012}.
To set the maximum number of iterations, we apply the Morozov principle; the
algorithm only proceeds as long as:
\begin{equation}
\chi^2= \frac{1}{2\,K} \left( \sum^K_{k=1} f_{A,k}\left(x_k\right) +  
\sum^K_{k'=1} f_{B,k'}([\M{H}\,\V{x}]_{k'}) \right) < 1\,.
\end{equation}
In our experiments, this criterion seems to stop the algorithm close to the
optimum, as  can be seen  in \Fig{fig:RMSDG1} and \Fig{fig:costDG0}.

From \Fig{fig:Alpha}, it can be seen that the parameter $\alpha$ has a strong
effect on the speed of convergence but has little influence on the quality.
However, if $\alpha$ is too large (\eg $\alpha=1$ in  \Fig{fig:Alpha}), the
steps are  too large and the criterion $\chi^2$ is well below $1$ even after the first iteration. As consequence, $\alpha$ is set such
that $\chi^2>1$ for the first few iterations.

Such an
automatic tuning works  only for the  Gaussian likelihood. In the  absence of
noise, for the Poisson likelihood and   the classical
projection, we select the number of iterations and $\alpha$ that maximize
$\SNRxb$.

\subsection{Gaussian Noise}
\label{sec:ExpGauss}
We first compare the classical projection with the proximity operator derived
from the Gaussian likelihood.
In the noiseless case, the proximity operator improves  $\SNRxb$ by
about $0.5\,$dB.
However, the visual differences between both reconstructions  
back-projected in the $z_0$ plane are barely noticeable 
as shown on Figures \ref{fig:CPs0}  and \ref{fig:GPs0}.

For the noisy scenario, the reconstruction error as a
function of the standard deviation of the noise is shown  in \Fig{fig:Gaussian}.
We observe that the use of the proximity operator always improves 
 $\SNRxb$ by at least $0.5\,$dB compared to the classical projection. 
%even in the absence of noise.
When the noise is
$\sigma=0.3$ or higher (\ie, the SNR of the measurements is lower than
$2.4\,$dB), the classical projection fails to properly estimate any phase. As
consequence, the twin image  appears much more clearly in the back-propagated field to $z_0$ in the  classical projection case
than with the proposed proximity operator, as can be seen in \Fig{fig:CPs3}
and \Fig{fig:GPs3}.

\subsection{Photon Counting}
To test the proximity operator derived for the Poisson likelihood we performed
simulations while varying the  illumination and without any background
emission ($b_k=0$), in which case the proximity operator  is given by
\Eq{eq:ph2nd}.
We compared its performance to that of the classical projection for an
illumination varying from $10^5$ to $10^9$ photons in each plane. 
Compared to the classical projection, the proposed
proximity operator always improves  $\SNRxb$, as  can be seen in
\Fig{fig:Photon}.   The performance gap with to classical projection becomes
 smaller as the number of photons increases.

\subsection{Low-Light Conditions}

\begin{figure*}[htbp] \noindent
\begin{minipage}[t]{.31\linewidth}\centering
\fbox{\includegraphics[height=\linewidth]{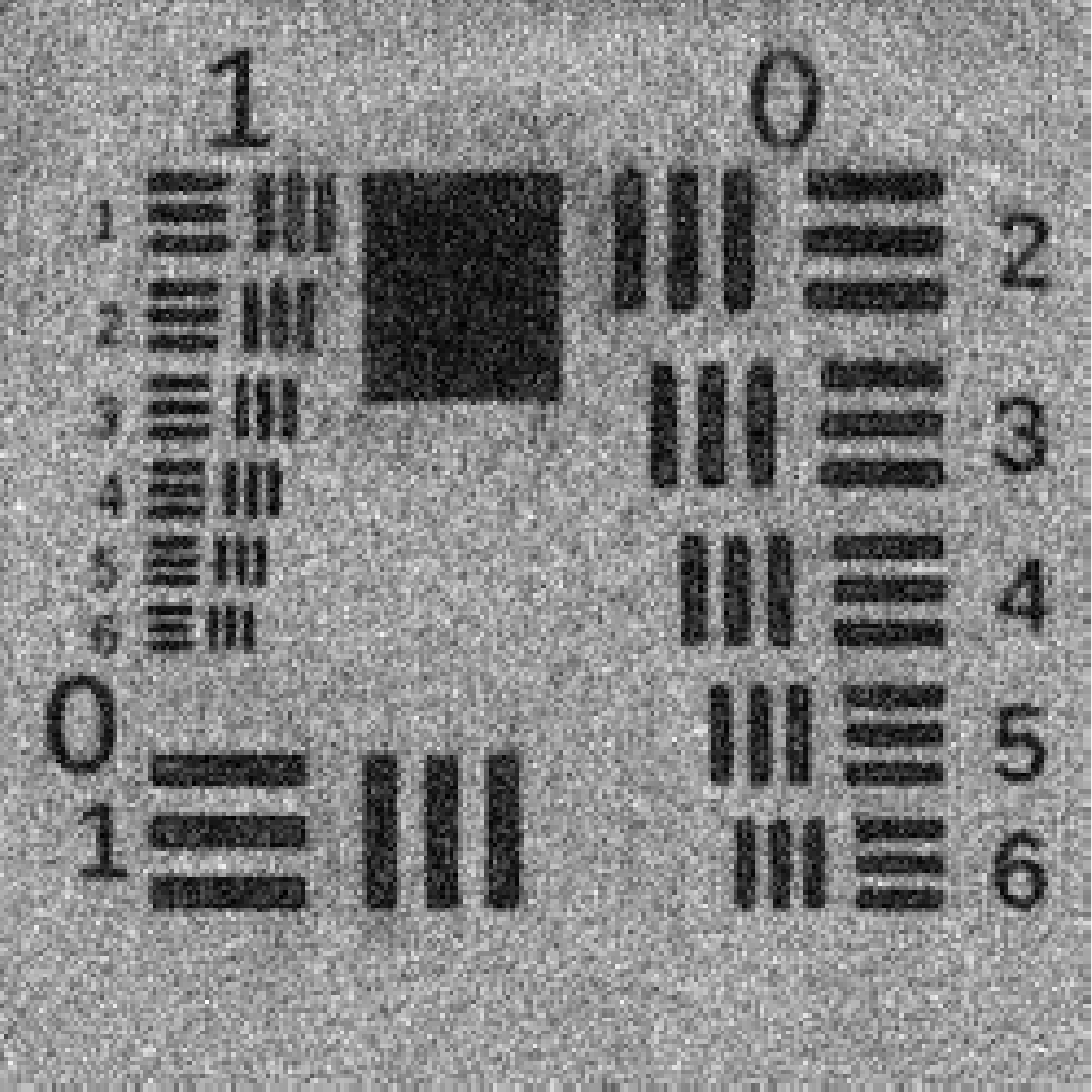}}
\caption{Central $250\times 250$ pixels of the test chart recovered from 8
planes measurements 
using the proximity operator for sum-of-intensity measurements and
back-propagated to $z=0$.}
\label{fig:SbSz8}
\end{minipage}\hfill
\begin{minipage}[t]{.31\linewidth}\centering
\fbox{\includegraphics[height=\linewidth]{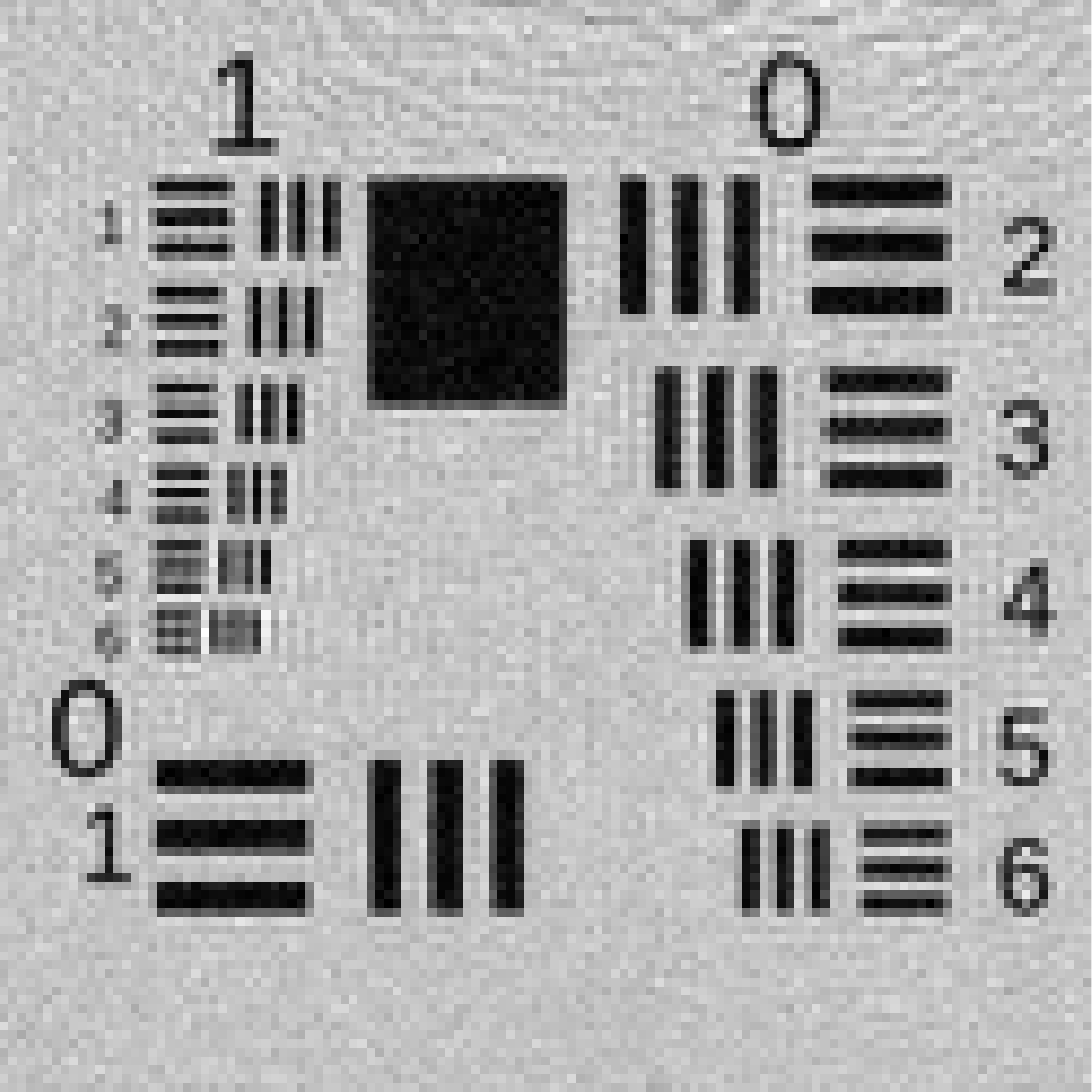}}
\caption{Central $125\times 125$ pixels of the test chart recovered from  8
planes measurements using the proximity operator presented in
Section~\ref{sec:Gauss} and back-propagated to $z=0$.}
\label{fig:NSbSz8}
\end{minipage}\hfill
\begin{minipage}[t]{.31\linewidth}\centering
\fbox{\includegraphics[height=\linewidth]{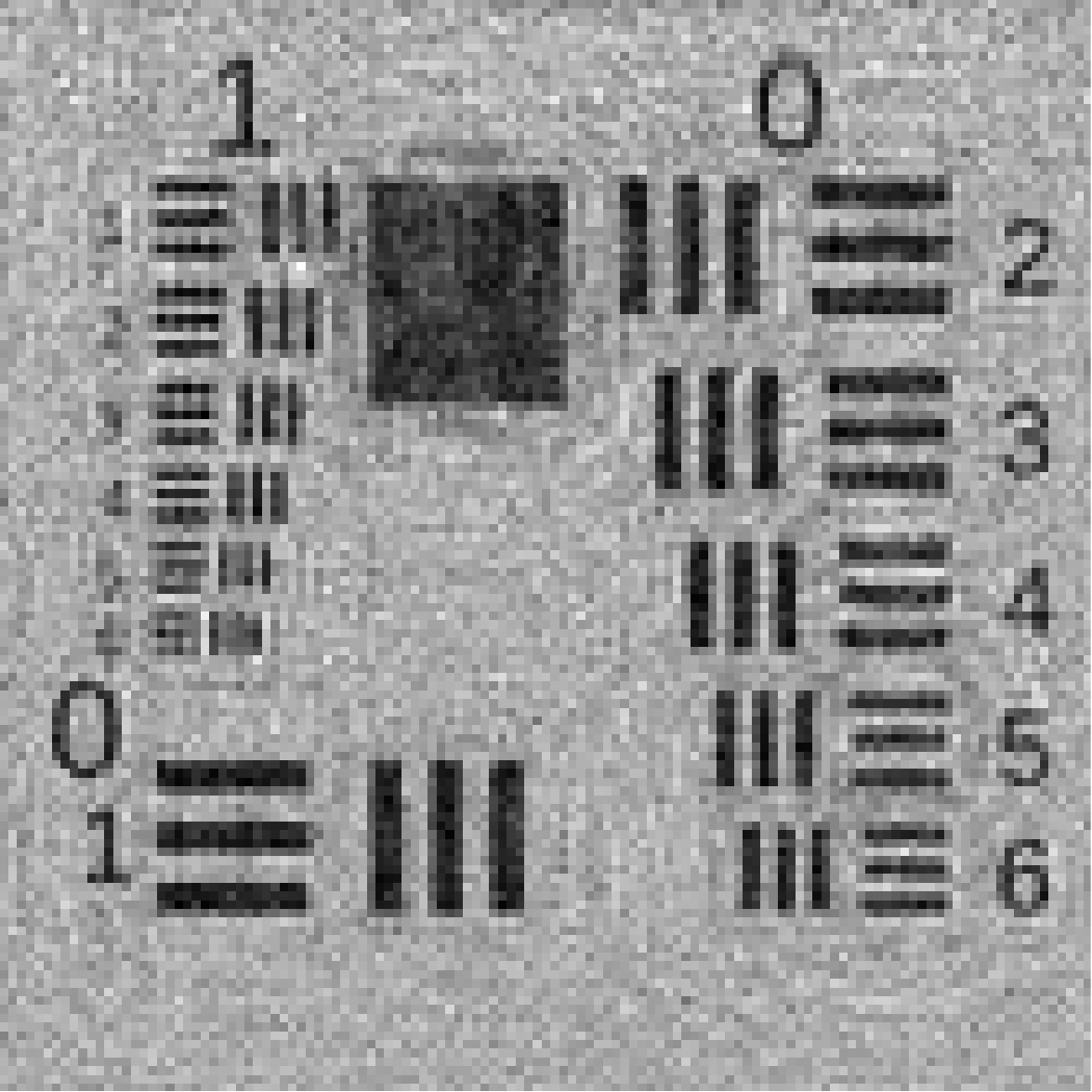}}
\caption{Central $125\times 125$ pixels of the test chart recovered from the 2
planes measurements using the proximity operator presented in
Section~\ref{sec:Gauss} and back-propagated to $z=0$.}
\label{fig:NSbSz2}
\end{minipage}
\end{figure*}

In low light, most  detection devices are plagued by dark current, which can be
modeled by an additive background emission $b_k > 0$.
For   illuminations from $10^5$ to
$10^9$ photons, we simulated the measured intensity $d_k$ at pixel $k$ following
a Poisson distribution $\mathcal{P}$, so that
\begin{equation}
\label{eq:poisson}
d_k = \mathcal{P}\left(  \Abs{x_k}^2  + b\right) \,,
\end{equation}
where the dark current was set to $b=3\,e^{-}$  per pixel. The
reconstruction SNR as a function of illumination is shown on \Fig{fig:LLRMS} for the classical
projection, the Poisson-likelihood proximity operator and the Gaussian-likelihood proximity
operator assuming a signal-dependent  Gaussian  noise with
mean $b=3$ and an inverse variance $w_k$ at pixel $k$ estimated as
\begin{equation}\label{eq:nonsta}
w_k =  1/\max(d_k,b)\,.
\end{equation}
In  \Fig{fig:LLRMS}, we see that the two proximity operators have  a very
similar performance and perform better
than the classical projection. Interestingly, both proximity operators. This
means that, even with a quite low dark current  (here $b=3$), the
approximation of a Poisson noise with the non-stationary Gaussian noise given 
in \Eq{eq:nonsta} is  good.

\subsection{Undersampled Fringes: Trading SNR for Resolution.}

We tested the  sum-of-intensity proximity operator derived in 
section~\ref{sec:SumProx} in the case where the fringes are not sufficiently sampled by the detector. Given the adequately
sampled complex amplitude $\V{g}_p \in\Complexes^{K_1 \times K_2 }$ in the
detector plane $z_p$, we simulated  $(2\times 2)$  subsampled intensity
measurements $\V{d}_p \in\Reals^{M_1 \times M_2 }$  with $K= 2\,M$ using the
 direct model
\begin{eqnarray}
\V{g}_{p}&=&\M{H}_p\cdot\V{r}\,,\\
d_{p,m_1, m_2}&=&\Abs{g_{p,2\,m_1, 2 m_2} }^2 + \Abs{g_{p, 2\,m_1 +1,
2m_2} }^2\nonumber	 \\ 
&& + \Abs{g_{p, 2\,m_1, 2\,m_2+1} }^2
+\Abs{g_{p,2\,m_1+1, 2\,m_2+1} }^2 + \V{n}_p\,, 
\end{eqnarray}
where $\M{H}_p$ is the propagation operator from the plane $z_1$ to $z_p$.
As in the previous experiments, we estimated the complex amplitude  
$\V{x}^\best$ in the plane $z_1$.

The strategy without regularization is only viable when there are  sufficiently
many measurements $(P\times M_1 \times M_2)$ as compared to the number of
unknowns $(2\times K_1 \times K_2 = 8 \times  M_1 \times M_2)$. To increase the
number of measurements, we modified  the proposed setup and estimated
$\V{x}^\best$ in the plane $z_1$ from $P=8$ measurements.

The maximum-likelihood solution in this case is given by
\begin{equation}
\label{eq:I-lkl4}
\V{x}^\best \in \argmin_{\V{x}\in\Complexes^K} \sum_{p=1}^{8}
\sum_{k=1}^{K} f_{p,k}\left(\left[\M{H}_p\cdot\V{x}\right]_{k}\right)\,.
\end{equation}
% 
% \begin{equation}
% \label{eq:I-lkl4}
% \V{x}^\best = \argmin_{\V{x}\in\Complexes^K} 
% \sum_{k=1}^{K} f_{A,k}\left(\V{x}_{k}\right) +
% \sum_{k=1}^{K} f_{B,k}\left(\left[\M{H}_B\cdot\V{x}\right]_{k}\right) +
% \sum_{k=1}^{K} f_{C,k}\left(\left[\M{H}_C\cdot\V{x}\right]_{k}\right) +
% \sum_{k=1}^{K} f_{D,k}\left(\left[\M{H}_D\cdot\V{x}\right]_{k}\right) \,,
% \end{equation}
%  
It is solved by  means of the PPXA
algorithm~\cite{PustelnikChauxPesquet2011PPXA}, which is a generalization of the
Douglas-Rachford algorithm that minimizes the sum of  more than two functions.  

We simulated  intensity measurements for eight planes 
  taken at $z_1
=1\,$cm, $z_2 =1.5\,$cm, $z_3 = 2\,$cm, $z_4 =2.5\,$cm, $z_5 = 3\,$cm, $z_6
=3.5\,$cm, $z_7 =4\,$cm and  $z_8 = 4.5\,$cm. These measurements were
corrupted with additive Gaussian noise of variance $\sigma=0.5$
(coresponding to $\SNR{}=-2.1\,$dB).

We have estimated the $1024\times968$ pixels complex amplitude $\V{x}^\best$ in
the plane $z_1$ from these eight $512\times 484$ pixels intensity measurements
using the proposed proximity operator for sum of intensities with $f$ derived for the
Gaussian likelihood (\ref{eq:Glkldef}). A zoom on the central part of the wave
back-propagated  to $z_0$ is presented in \Fig{fig:SbSz8}. It illustrates the
effectiveness of the proposed proximity operator to recover fine details and
increase the resolution.
This can be compared with two reconstructions without superresolution using the
same PPXA algorithm but with the proximity operator derived in
Section~\ref{sec:Gauss}. One,  shown on   \Fig{fig:NSbSz8}, was done with the
same measurements ( $8 \times M_1 \times M_2$ measurements for $2 \times  M_1 \times
M_2$ unknowns). The other,  shown on   \Fig{fig:NSbSz2}, is using only the
measurements in the two planes $z_1$ and $z_2$ to get  the same number
measurements than unknowns ($2 \times  M_1 \times M_2$).
% To have a similar conditioning of the reconstruction problem (\ie{} to keep
% the same number of unknown as number of measurements), this reconstruction
% without super-resolution is obtained   using the first two ($z_A$ and $z_B$)
% images obtained by the DR algorithm used in section~\ref{sec:ExpGauss}.
Compared to these non-superresolved reconstructions,  the resolution improvement
is obvious. However, this improvement is acquired at the cost of a moderate
increase in noise compared to the reconstruction shown on the \Fig{fig:NSbSz8}.
Indeed, the non-superresolved reconstruction appears less noisy as the ratio of the number of unknowns over the number of measurements is
more favorable. This reconstruction noise is similar to  in the
non-superresolved reconstruction using only two planes to get  the same number
measurements than unknowns  shown on the \Fig{fig:NSbSz2}.
%  \subsection{Laplace noise} Does not work

%
% \subsection{Low light noise}
% As it is separable, the proposed proximity operator can cope
% with non stationary noise. We thus studied its performance in a low light regime
% where the photon noise can be approximated by a Gaussian non stationnary
% noise. We simulate the noisy measurement $d_k$ as:
% \begin{equation}
% \label{eq:poisson}
% d_k = \textrm{round}\left(\gamma\,\mathcal{P}\left(  \Abs{x_k}^2 \right) +
% n_k\right)\,,
% \end{equation}
% where $\gamma$ is the reciprocal of the quantification level, $\mathcal{P}()$
% the Poisson distribution and $n_k$ additive Gaussian noise of variance $\sigma^2$ (\eg{}
% read out noise).
%
% For the reconstruction, the inverse variance $w_k$ at pixel $k$ is estimated
% as:
% \begin{equation}
%  w_k = \left(\gamma \max(d_k,0) + \sigma^2 + 1/12 \right)^{-1}\,,
%  \end{equation}
% where $1/12 = \int_{-1/2}^{1/2} x^2 dx$ account for the quantization noise.
%

\section{Conclusion}
We considered the problem  of the phase retrieval from noisy intensity
measurements.
From the maximum-likelihood formulation, we derived  proximal operators for
intensity measurements corrupted with Gaussian noise or Poisson noise. We
further expanded these proximity operators for cases where fringes are not
properly sampled. When plugged into the Gerchberg-Saxton algorithm in place of
the classical projection, it showed superior results.  As it can be plugged into
any projection-based algorithm, it can provide an improvement of the  performance
for many phase-retrieval algorithms without changing the core of the
optimization procedure.

\section*{Acknowledgements}
This work is supported by the Sinergia project “Euclid: precision cosmology in
the dark sector" from the Swiss National Science Foundation and by the French ANR POLCA project (Processing of
pOLychromatic interferometriC data for Astrophysics, ANR-10-BLAN-0511).

% Bibliography
\bibliography{ProxPaper}

% Full bibliography added automatically for Optics Letters submissions
% Note that this extra page will not count against page length
%\ifthenelse{\equal{\journalref}{ol}}{\clearpage
%\bibliographyfullrefs{ProxPaper}
%}{}

%Manual citation list
% \begin{thebibliography}{1}
% \bibitem{Zhang:14}
% Y.~Zhang, S.~Qiao, L.~Sun, Q.~W. Shi, W.~Huang, %L.~Li, and Z.~Yang,
%   \enquote{Photoinduced active terahertz metamaterials with nanostructured
%   vanadium dioxide film deposited by sol-gel method,} Opt. Express \textbf{22},
%   11070--11078 (2014).
% \end{thebibliography}

\end{document}